\documentclass{aa}
\usepackage{lscape}
\usepackage{graphicx}
\usepackage{txfonts}
\usepackage{longtable}
\usepackage{natbib}
\usepackage{url}
\usepackage{float}
\usepackage{aalongtable}
\usepackage{ulem}
\usepackage{color}
\usepackage{soulutf8}
\usepackage{textcomp}
\usepackage{hyperref}

\begin{document}

\title{The CARMENES search for exoplanets around M dwarfs}  
\subtitle{A deep learning approach to determine fundamental parameters of target stars}

\titlerunning{A deep learning approach for fundamental parameters}
\author{V.\,M.~Passegger\inst{1,2}
        \and
        A.~Bello-Garc\'ia\inst{3}
        \and
        J.~Ordieres-Mer\'e\inst{4}
	    \and
        J.\,A.~Caballero\inst{5}
        \and
        A.~Schweitzer\inst{1}
        \and 
        A.~Gonz\'alez-Marcos\inst{6}
        \and 
        I.~Ribas\inst{7,8}
        \and 
	    A.~Reiners\inst{9}
        \and
        A.~Quirrenbach\inst{10}
        \and
        P.\,J.~Amado\inst{11}
        \and
        M.~Azzaro\inst{12}
        \and
        F.\,F.~Bauer\inst{11}
        \and
        V.\,J.\,S.~B\'ejar\inst{13,14}
        \and
        M.~Cort\'es-Contreras\inst{5}
        \and 
        S.~Dreizler\inst{9}
        \and
        A.\,P.~Hatzes\inst{15}
        \and
        Th.~Henning\inst{16}
        \and
        S.\,V.~Jeffers\inst{9}
        \and 
        A.~Kaminski\inst{10}
        \and 
        M.\,K\"urster\inst{16}
        \and 
        M.~Lafarga\inst{7,8}
        \and
        E.~Marfil\inst{17}
        \and
        D.~Montes\inst{17}
        \and
        J.\,C.~Morales\inst{7,8}
        \and 
        E.~Nagel\inst{15,1}
        \and
        L.\,M.~Sarro\inst{18}
        \and
        E.~Solano\inst{5}
        \and
        H.\,M.~Tabernero\inst{19}
        \and
        M.~Zechmeister\inst{9}
        }

\institute{
        Hamburger Sternwarte, Gojenbergsweg 112, D-21029 Hamburg, Germany \newline
        \email{vpassegger@hs.uni-hamburg.de}
        \and
        Homer L. Dodge Department of Physics and Astronomy, University of Oklahoma, 440 West Brooks Street, Norman, OK 73019, United States of America
        \and
        Departamento de Construcci\'on e Ingenier\'ia de Fabricaci\'on, Universidad de Oviedo, c/ Pedro Puig Adam, Sede Departamental Oeste, M\'odulo 7, 1$^a$ planta, E-33203 Gij\'on, Spain
        \and
        Departamento de Ingenier\'ia de Organizaci\'on, Administraci\'on de Empresas y Estad\'istica, Universidad Polit\'ecnica de Madrid, c/ Jos\'e Guti\'errez Abascal 2, E-28006 Madrid, Spain
	    \and
        Centro de Astrobiolog\'ia (CSIC-INTA), ESAC, Camino bajo del castillo s/n, E-28692 Villanueva de la Ca\~nada, Madrid, Spain
        \and
        Departamento de Ingenier\'ia Mec\'anica. Universidad de la Rioja. San Jos\'e de Calazanz 31, 26004 Logro\~no, La Rioja, Spain
        \and
        Institut de Ci\`encies de l'Espai (CSIC-IEEC), Campus UAB, c/ de Can Magrans s/n, E-08193 Bellaterra, Barcelona, Spain
        \and
        Institut d'Estudis Espacials de Catalunya (IEEC), E-08034 Barcelona, Spain
        \and
        Institut f\"ur Astrophysik, Georg-August-Universit\"at, Friedrich-Hund-Platz 1, D-37077 G\"ottingen, Germany
        \and
        Landessternwarte, Zentrum f\"ur Astronomie der Universt\"at Heidelberg, K\"onigstuhl 12, D-69117 Heidelberg, Germany
        \and
        Instituto de Astrof\'isica de Andaluc\'ia (IAA-CSIC), Glorieta de la Astronom\'ia s/n, E-18008 Granada, Spain
        \and
        Centro Astron\'omico Hispano-Alem\'an (CSIC-MPG), Observatorio Astron\'omico de Calar Alto,  Sierra de los Filabres, E-04550 G\'ergal, Almer\'ia, Spain
        \and
        Instituto de Astrof\'{\i}sica de Canarias, c/ V\'ia L\'actea s/n, E-38205 La Laguna, Tenerife, Spain
        \and
        Departamento de Astrof\'{\i}sica, Universidad de La Laguna, E-38206 La Laguna, Tenerife, Spain
        \and
        Th\"uringer Landessternwarte Tautenburg, Sternwarte 5, D-07778 Tautenburg, Germany
        \and        
        Max-Planck-Institut f\"ur Astronomie, K\"onigstuhl 17, D-69117 Heidelberg, Germany
        \and
        Departamento de F\'{\i}sica de la Tierra y Astrof\'{\i}sica and IPARCOS-UCM (Instituto de F\'{\i}sica de Part\'{\i}culas y del Cosmos de la UCM), Facultad de Ciencias F\'{\i}sicas, Universidad Complutense de Madrid, E-28040 Madrid, Spain
        \and
        Depto. de Inteligencia Artificial, UNED, Juan del Rosal, 16, Madrid, 28040, Spain
        \and
        Instituto de Astrof\'isica e Ci\^encias do Espa\c{c}o, Universidade do Porto, CAUP, Rua das Estrelas, 4150-762 Porto, Portugal
        }

\date{Received 30 June 2020 / Accepted 28 July 2020}

\abstract
{Existing and upcoming instrumentation is collecting large amounts of astrophysical data, which require efficient and fast analysis techniques. 
We present a deep neural network architecture to analyze high-resolution stellar spectra and predict stellar parameters such as effective temperature, surface gravity, metallicity, and rotational velocity. 
With this study, we firstly demonstrate the capability of deep neural networks to precisely recover stellar parameters from a synthetic training set. Secondly, we analyze the application of this method to observed spectra and the impact of the synthetic gap (i.e., the difference between observed and synthetic spectra) on the estimation of stellar parameters, their errors, and their precision.
Our convolutional network is trained on synthetic PHOENIX-ACES spectra in different optical and near-infrared wavelength regions. 
For each of the four stellar parameters, $T_{\rm eff}$, $\log{g}$, [M/H], and $v \sin{i}$, we constructed a neural network model to estimate each parameter independently. 
We then applied this method to 50 M dwarfs with high-resolution spectra taken with CARMENES (Calar Alto high-Resolution search for M dwarfs with Exo-earths with Near-infrared and optical \'Echelle Spectrographs), which operates in the visible (520--960\,nm) and near-infrared wavelength range (960--1710\,nm) simultaneously. 
Our results are compared with literature values for these stars. 
They show mostly good agreement within the errors, but also exhibit large deviations in some cases, especially for [M/H], pointing out the importance of a better understanding of the synthetic gap. }
%
\maketitle
%

\section{Introduction}
\label{introduction}



The determination of stellar parameters in M dwarfs has always been challenging. M dwarfs are smaller, cooler, and fainter than Sun-like stars. Due to their faintness, higher stellar activity with sometimes strong magnetic fields, stronger line blends, and the lack of true continuum, well-established photometric and spectroscopic methods are brought to their limits. 
In the literature there are several methods to estimate M-dwarf parameters such as effective temperature $T_{\rm eff}$, surface gravity $\log{g}$, and metallicity.
%
In particular, the measurement of spectral line pseudo-equivalent widths (pEWs) is a widely used method for metallicity determination. \cite{Neves2013,Neves2014} measured pEWs for the HARPS guaranteed time observations of M dwarfs and calibrated them with photometric relations from \cite{Neves2012} for metallicity and from \cite{Casagrande2008} for effective temperature. 
Metallicity relations based on equivalent widths were derived by \cite{Newton2014} using low-resolution spectra ($R \sim$ 2000) in the $J$, $H$, and $K$ bands. 
They calibrated their method with binary systems with an F-, G-, or K-type primary and an M-dwarf secondary. Pseudo-EWs in the 
$K$- and $H$-band and H$_2$O indices were measured by \cite{Khata2020} to determine metallicities and effective temperatures as well as radii, luminosities, spectral types, and absolute magnitudes for 53 M dwarfs. 
Combining pEWs with empirical calibrations and spectral synthesis, \cite{Veyette2017} derived $T_{\rm eff}$ and Ti and Fe abundances for 29 M dwarfs with high-resolution $Y$-band spectra. 

A method considered to be very precise is the calibration with M dwarfs that have an F, G, or K binary companion with known metallicity. Many of the above mentioned relations were calibrated using FGK+M binary systems \citep[e.g.,][]{Newton2014}. \cite{Mann2013a} identified metallicity sensitive features in low-resolution visible, $J$-, $H$-, and $K$-band spectra of 112 late-K to mid-M dwarfs in binary systems with higher mass companions, from which they derived different metallicity calibrations. The same relations were used by \cite{Rodriguez2019} to determine metallicities from mid-resolution $K$-band spectra for 35 M dwarfs of the {\it K2} mission. Other photometric calibrations using FGK+M binary systems are presented by \cite{Bonfils2005}, \cite{Casagrande2008}, \cite{JohnsonApps2009}, \cite{SchlaufmanLaughlin2010}, and \cite{Neves2012}, among others. Several spectroscopic calibrations can be found in \cite{Rojas-Ayala2010}, \cite{Dhital2012},  \cite{Terrien2012}, \cite{Mann2014}, \cite{Mann2015}, and \cite{Montes2018}.

Another approach to determine stellar parameters for M dwarfs is the calculation of different kinds of spectral indices. Many of these indices were calibrated using FGK+M binaries, as mentioned above.
\cite{GaidosMann2014} derived $T_{\rm eff}$ from $K$-band M-dwarf spectra by calculating spectral curvature indices. For metallicity they used relations of atomic line strength based on \cite{Mann2013a}. 
Also working in the $K$ band, \cite{RojasAyala2012} calculated the H$_2$O-K2 index, quantifying the absorption from H$_2$O opacity, to determine $T_{\rm eff}$ from low-resolution ($R \sim$ 2700) M-dwarf spectra. 
\cite{Newton2015} measured 13 spectral indices and 26 pEWs in the near-infrared $H$ band to estimate metallicities of the MEarth transiting planet survey and the cool {\it Kepler} objects of interest. For temperature determination they employed the H$_2$O-K2 index from \citet{RojasAyala2012} and spectral indices from \cite{Mann2013b}. 
Molecular indices of CaH and TiO were calculated by \cite{WoolfWallerstein2006} to derive a relation between those indices and metallicities for 76 late-K and M dwarfs. To test their results, they compared measurements of several M dwarfs with higher mass companions to the metallicities of the primaries and find good agreement. 
\cite{Johnson2012} combined several existing photometric relations to derive the stellar properties of KOI-254.

Stellar parameters can also be derived from interferometric measurements. However, only a limited number of stars is accessible for such observations, since they have to be bright and nearby.  
\cite{Boyajian2012} present interferometric angular diameters for 26 K and M dwarfs measured with the CHARA array and for seven K and M dwarfs from the literature. With parallaxes and bolometric fluxes they computed absolute luminosities, radii, and $T_{\rm eff}$. They also calculated empirical relations for K0 to M4 dwarfs to connect $T_{\rm eff}$, radius, and luminosity to broadband color indices and [Fe/H].
\cite{Maldonado2015} estimated $T_{\rm eff}$ from pEWs calibrated with interferometric temperatures from \cite{Boyajian2012} and metallicities from pEWs calibrated with the \cite{Neves2012} relations. They constructed a mass-radius relation using interferometric radii \citep{Boyajian2012,vonBraun2014} and masses from eclipsing binaries \citep{Hartman2015}. From this they calculated surface gravities, $\log{g}$. 
Other works that derived $T_{\rm eff}$ from angular diameters include, for example, \cite{Segransan2003}, \cite{Demory2009}, \cite{vonBraun2014}, and \cite{Newton2015}. Of them, \cite{Segransan2003} also determined $\log{g}$ from their measured masses and radii.


A not very commonly used method for stellar parameter determination is the principle component analysis. \cite{Paletou2015} and \cite{He2019} show that this method could in principle be used to derive effective temperature for K- and M-type stars and abundances for dwarfs and giants, respectively.

With the improvement of synthetic models for cool stellar atmospheres, model fits to low- or high-resolution spectra are getting more powerful. Several model sets are based on the stellar atmosphere code PHOENIX \citep{Hauschildt1992,Hauschildt1993}. 
The BT-Settl models \citep{Allard2012,Allard2013} were used by \cite{Veyette2017} and \cite{Rajpurohit2018}, who determined $T_{\rm eff}$, $\log{g}$, and metallicities for M dwarfs from CARMENES high-resolution spectra \citep{Reiners2018a}. 
\cite{GaidosMann2014} and \cite{Mann2015} also derived $T_{\rm eff}$ of M dwarfs from fitting BT-Settl models, but to low-resolution visible SNIFS spectra. 
\cite{Kuznetsov2019} determined $T_{\rm eff}$, $\log{g}$, [Fe/H], and $v \sin{i}$ of 153 M dwarfs by fitting mid-resolution visible spectra from X-shooter at VLT using BT-Settl models.

\cite{ZborilByrne1998} \citep[see also][]{Zboril1998} fitted PHOENIX models \citep{AllardHauschildt1995} to derive $\log{g}$ and [M/H] for 11 M dwarfs with high-resolution visible spectra. To estimate $T_{\rm eff}$ they used photometric indices. 
\cite{Bean2006} generated synthetic spectra based on the PHOENIX NextGen models \citep{Hauschildt1999} using the stellar analysis code MOOG \citep{Sneden1973} to determine $T_{\rm eff}$ and [M/H] for three planet-hosting M dwarfs. 
After several recent updates of line lists and the equation of state, the PHOENIX models were used by \cite{Passegger2018} \citep[PHOENIX-ACES, see][]{Husser2013}, \cite{Passegger2019}, and \cite{Schweitzer2019} \citep[both using the SESAM equation of state, see][]{Meyer2017} to derive $T_{\rm eff}$, $\log{g}$, and [Fe/H] of M dwarfs observed with CARMENES in the visible and near-infrared wavelength ranges. 
\cite{Birky2017} determined the same parameters for late-M and early-L dwarfs from fitting PHOENIX models to high-resolution near-infrared APOGEE spectra ($R \sim$ 22\,500). 
Another source for synthetic models widely used are the MARCS model atmospheres \citep{Gustafsson2008}. \cite{Onehag2012} used MARCS atmospheres together with the {\tt SME} package \citep{ValentiPiskunov1996,ValentiFischer2005} to calculate synthetic models with specific parameters on the fly and to fit them to high-resolution $J$-band spectra of eight M dwarfs. \cite{Souto2017} generated synthetic spectra using the {\tt Turbospectrum} code \citep{AlvarezPlez1998,Plez2012} and MARCS atmospheres and fitted them to high-resolution APOGEE spectra of two M-dwarf exoplanet hosts to determine 13 element abundances. 
\cite{Souto2018} employed MARCS and BT-Settl model atmospheres \citep{Allard2013} together with {\tt Turbspectrum} to find $T_{\rm eff}$, $\log{g}$, and eight element abundances of the exoplanet host \object{Ross~128} (M4.0\,V). Recently, \cite{Souto2020} used the {\tt Turbospectrum} code and MARCS atmospheres with updated APOGEE atomic and molecular line lists \citep{Shetrone2015} to calculate synthetic spectra and fitted them to 21 M-dwarf high-resolution $H$-band spectra observed with APOGEE. 

In recent years, machine learning (ML) has proved to be a powerful tool in many fields and also found its way into astrophysics and stellar parameter determination. 
Machine learning techniques are used in several areas of astrophysics, such as galaxy morphology prediction \citep{Dieleman2015} and classification \citep{Wu2018}, detection of bar structures in galaxy images \citep{Abraham2018}, determining the evolutionary states of red giants from asteroseismology \citep{Hon2017}, and classifying variable stars from analysis of light curves \citep{Mahabal2017}. Early applications of neural networks to characterize stellar spectra can be found in \cite{vonHippel1994}, \cite{Gulati1994}, and \cite{Singh1998}, for example. \cite{Bailer-Jones1997} trained an artificial neural network with synthetic spectra to determine $T_{\rm eff}$, $\log{g}$, and [M/H] for over 5000 stars of spectral types B to K. 

\cite{Sharma2019} compared different ML algorithms, such as artificial neural networks (ANN), random forests, and convolutional neural networks, to classify stellar spectra. They report that their convolutional neural network achieved better accuracy than the other ML algorithms, and point out the importance of a sufficiently large training set. 
\cite{Sarro2018} used genetic algorithms for selecting features such as equivalent widths and integrated flux ratios from BT-Settl model atmospheres. They estimated $T_{\rm eff}$, $\log{g}$, and [M/H] for M dwarfs with eight different regression models and ML techniques, and compared the results to classical $\chi^2$ and independent component analysis coefficients. 
\cite{Whitten2019} present the Stellar Photometric Index Network Explorer ({\tt SPHINX}), an artificial neural network approach to estimate $T_{\rm eff}$ and [Fe/H] from J-PLUS broad- and intermediate-band optical photometry, and synthetic magnitudes. {\tt SPHINX} is able to successfully estimate temperatures and metallicities for stars in the range $4500\,{\rm K} < T_{\rm eff} < 8500$\,K and down to [Fe/H] $\sim -3$.

{\tt The Cannon} \citep{Ness2015} and {\tt The Cannon2} \citep{Casey2016} were designed to derive stellar parameters from APOGEE spectra. {\tt The Cannon} is a data-driven approach that is trained with observed spectra with known parameters from the APOGEE pipeline. For training they used only 542 reference stars. They show that {\tt The Cannon} is able to provide accurate $T_{\rm eff}$, $\log{g}$, and [Fe/H] for all 55\,000 stars of APOGEE DR10, even for those with low signal-to-noise ratio (S/N) around 50.
\cite{Birky2020} also used {\tt The Cannon} and determined $T_{\rm eff}$, $\log{g}$, [Fe/H], and detailed abundances for 5875 M dwarfs from the APOGEE and $Gaia$ DR2 surveys. 

\cite{Fabbro2018} trained the convolutional neural network {\tt StarNet} on observed APOGEE spectra as well as synthetic MARCS and ATLAS9 model atmospheres. They applied {\tt StarNet} to 148\,724 and 21\,787 stars, respectively, with temperatures between 4000\,K and 5000\,K, and measured $T_{\rm eff}$, $\log{g}$, and [Fe/H]. They find that {\tt StarNet} is capable of deriving parameters close to those determined from the APOGEE pipeline when trained with observed spectra, although there are some larger differences for lower metallicities and higher temperatures. For {\tt StarNet} trained on synthetic spectra the intrinsic error is about twice as large than for observed spectra, although the resulting parameters are very similar. Additionally, \cite{Fabbro2018} gave a detailed description on neural networks. 
\cite{Leung2019} also derived stellar parameters for the whole APOGEE dataset using the Bayesian neural network {\tt astroNN}, which was trained with observed spectra from the APOGEE pipeline over the full spectral range. Additionally, they determined 19 individual element abundances from specific wavelength ranges using mini-networks. 

Recently, \cite{Antoniadis2020} present their ML tool {\tt ODUSSEAS}, which is based on measuring pEWs of more than 4000 absorption lines in the optical. They trained their neural network with a set of HARPS spectra consisting of only 45 training and 20 test spectra. {\tt ODUSSEAS} can be applied to spectra with different resolutions because the tool adjusts the resolution of the HARPS spectra in the training step. Thanks to this capability, \cite{Antoniadis2020} successfully derived $T_{\rm eff}$ and [Fe/H] for several M dwarfs observed with different instruments. 

In the study presented here we follow two main interests. First of all, we are interested in providing insights into the capability of deep learning (DL, Section~\ref{sect:depp.learning}) to create models able to learn stellar parameters when different configurations (i.e., architectures, wavelength windows, combinations of stellar parameters) are considered. Despite the regularly found astrophysical applications of DL models \citep{li2017parameterizing,Fabbro2018,shallue2018identifying}, we want to gain better understanding of the effects that different DL architectures have in the model creation, as well as to understand the significance of different spectral windows adopted to create models.
The second interest is about considering the uncertainty induced when the training process is carried out on synthetic spectra and parameter estimation is made for observed spectra, which is not the common way of using DL models. To do this we applied the DL method to a test sample of 50 M dwarfs with high-resolution and high-S/N spectra to demonstrate the applicability of ANNs, trained with synthetic PHOENIX-ACES models \citep{Husser2013}, to observed spectra. 


In Section~\ref{Method} we explain the DL procedure and our ANN architecture. 
Section~\ref{Analysis} describes the PHOENIX-ACES synthetic model grid that we used for training the neural network, our strategy of spectrum preparation, the stellar sample, and the application of our neural network. The derived stellar parameters are presented in Section~\ref{Results}, together with a literature comparison and discussion. Finally, in Section~\ref{Summary} we give a short summary of this work. 


\section{Method}
\label{Method}

\subsection{Deep learning} 
\label{sect:depp.learning}

\begin{figure*}[!ht]
\centering
\includegraphics[width=0.9\textwidth]{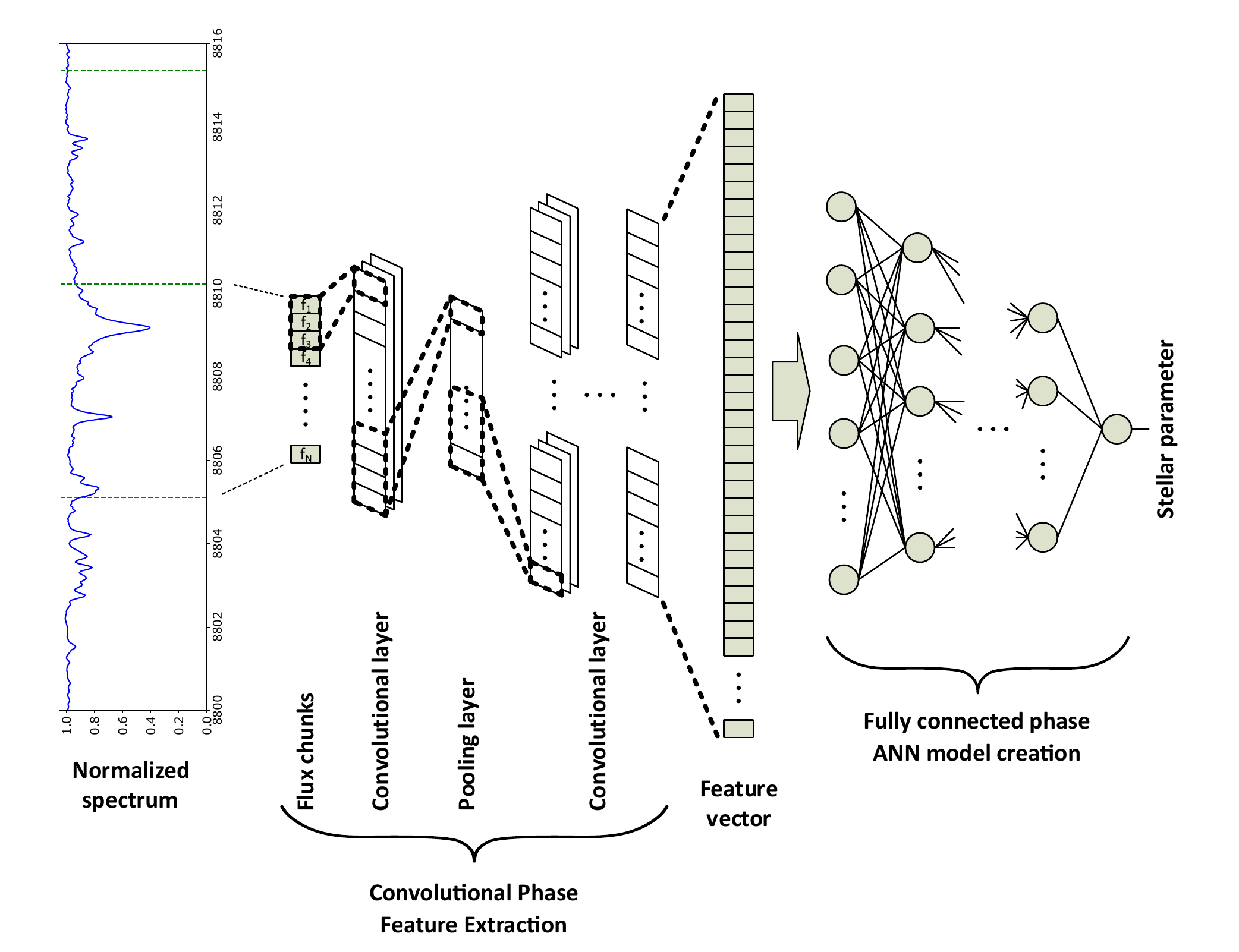}
\caption{Generic architecture for DL models in this work.}
\label{fig:NN}
\end{figure*}

Artificial intelligence is a broad area of computer science that develops systems able to perform tasks that are regularly seen to require human intelligence~\citep{mccarthy1981some, steels1993artificial, GonzalezMarcos2013}. 
Machine learning is an artificial intelligence technique looking to develop algorithms that can be ``taught'' how to learn patterns from data, instead of just transforming them. 
The ANN stands for a family of ML methods aiming to learn from data in a way inspired by the human brain structure \citep{zhang1998forecasting, anthony2009neural, Gong2016}. An ANN is constructed from different layers that are composed of a collection of artificial neurons or nodes. A node is characterized as a linear function of its input signals with weights. It is activated by a nonlinear activation function that can adopt different expressions.
Some of the most commonly used functions are linear, rectified linear unit (ReLU), sigmoid, or softmax.

Deep learning is a subset of ML methods that enables computational models consisting of multiple processing layers to learn representations of data with multiple levels of abstraction \citep{lecun2015deep, schmidhuber2015deep, Zheng2018}.
The main difference between traditional ML and DL techniques is that in ML the feature identification needs to be established by the user, whereas in DL the features are explored in an automatic way by means of different techniques, such as convolutional transformations of the input space data.

\subsection{ANN architectures}

The DL techniques employed in this paper utilize two different processing units to carry out the process. Firstly, a convolutional block creates feature sets starting from a single wavelength range of the one-dimensional spectrum, in this case a synthetic model created with known stellar parameters, and it ends with a large set of created features. Next, an ANN block implements a regressive modeling approach against the stellar parameter of interest (see Figure~\ref{fig:NN}). 

The ANN is built of neuron nodes organized by layers where every node is fully connected together with a feed-forward sequence, meaning that every output signal of the previous layer is connected to each node of the following layer.
Regarding the layer structure of the ANN, its input layer is the set of created features from the previous convolutional block, and the output layer is the predicted stellar parameter. 
In between those layers, there are several hidden layers consisting of neurons.
The more hidden layers the ANN contains, the ``deeper'' it is.
This structural configuration of DL models enables the processing of data with a nonlinear approach, not only because of the convolutional nature of its preprocessing steps, but also because the activation functions in the neurons of these ANNs are commonly non linear, such as the ReLU~\citep{petersen2018optimal}.

Regarding the convolutional block, the input layer is the one-dimensional synthetic spectrum, and is followed by several convolutional layers. In these layers the weights are applied as multipliers with tunable coefficients. 
When considered as a matrix operation, they can be seen as a filter to the input signal (the normalized flux values of the spectrum). These filters scan across the previous layer and convolve a certain section of the input with the weights to extract features from the previous layer. 
Later on, the ANN learns which features are the most relevant, and it correspondingly adjusts the filter coefficients. 
During or just after the convolutional layers a pooling layer can be included. The function of pooling layers is to progressively reduce the spatial size of the representation to decrease the amount of free parameters and computation in the network. 
It can be imagined as a window sliding across the previous convolutional layer, which calculates a function value of each sub-region (e.g., its average, its maximum, etc.). By selecting the maximum value, the designer of the DL architecture extracts the strongest overall features from the convolutional layer. In this case the layer is called max pooling layer. 
A structural schematic is presented in Fig.~\ref{fig:NN}.

\subsection{Training}

The DL model has a number of tunable parameters. In the following we refer to them as model parameters, in contrast to stellar parameters, which are the output of the DL model. 
These model parameters include number and length of filters in each convolutional layer, filter coefficients, pooling window size, and number and weights of connection nodes in the fully connected layers. At the beginning of the training phase the weights and coefficients are randomly set and get improved during the training. 
The training requires a reference set, which consists of a large number of observed or, in our case, synthetic spectra with known stellar parameters.
Although there is not an explicit formula linking the number of training samples and the error quantifying the performance of the model, still such a relationship does exist. The higher the number of training samples, the easier to get models with acceptable mean square error (MSE) and the more accurate the final stellar parameters will be. The way to increase the number of such training dataset is to reduce the parameter's step size in the training grid, for example by interpolation (see Section~\ref{Analysis.grid}). A limitation on the maximum number of training samples is given by the computing feasibility according to the available resources (GP-GPU memory and data transfer throughput). Also, since the DL is able to interpolate between the synthetic grid points to some extent, minimizing the step size would not add any additional value to the derived parameters.

The reference set is divided into a training set (95\,\%) and a validation set (5\,\%). 
During the training process the whole training set is fed into the DL model, all weights and filters are applied to the input flux in each layer, and stellar parameters are predicted. 
The obtained output depends on the model parameters but also on the DL architecture, such as the number of convolutions, pooling layers, the number of layers of the ANN, and the chosen type of activation functions. 
After the whole training set is processed, the output stellar parameters are compared to the known input stellar parameters, and the training error is estimated. 
This error is used to modify all the DL model parameters through a backward propagation using a gradient-based algorithm and a new training cycle starts. Each of those cycles are called an  ``epoch.'' 

Additionally, after each epoch the whole validation set is sent through the DL model to determine the validation error, which is estimated to be the MSE. In this case the background idea is not to correct the error, but to have an independent estimation of the error to be measured through epochs and to avoid over-fitting of the training dataset.
Over-fitting happens when the DL model describes the random variations in the dataset (e.g., tiny molecular features that appear as noise to the DL model) instead of the relations between variables (here, the stellar parameters). In this way the DL model moves from a regression into a memory tool.
Obviously, this evolution negatively impacts the ability of the model to generalize to new data. 
With the validation set, the learned DL model is evaluated regarding its performance on unseen data.
We created a regression approach able to estimate stellar parameters and not a system to identify or classify the training dataset like a memory.
Keeping the validation error consistently decreasing indicates that the adjustment of weights and coefficients progresses in the right direction to improve the DL model. 

The training continues until the minimum of the validation error is reached and, then, all weights and coefficients are fixed. It was commonly found after around 15 epochs. 
However, to be conservative, the algorithm usually ran between 35 and 50 epochs.
The architecture, that is the number of layers and sequence of convolutions, was provided before the training starts and was kept fixed during the process.

\subsection{Testing}

The trained DL model is then applied to the test set, which in this case was a randomly generated set of 100 synthetic spectra, not related to the reference set (i.e., never seen before, neither for training nor for validation). From this set, which is preserved during all the experiments, the quality of the created models is measured through the test error.
If the DL model performs well, which means that the average of the test error is lower than an adopted threshold depending on the stellar parameter under consideration, the training phase is considered complete and the DL model is applied to predict stellar parameters in observed spectra. 

In our particular case, for each stellar parameter, an individual DL model was built to predict $T_{\rm eff}$, $\log{g}$, [M/H], and $v \sin{i}$ separately, although several experiments were also conducted for predicting several parameters from the same model. The analysis of the architectures for each of these models is presented in Section~\ref{Results.Architecture}.

\section{Analysis}
\label{Analysis}

\subsection{Observational sample}
\label{Analysis.Sample}

To test our DL method we used the same template spectra as in \cite{Passegger2019} (in the following referred to as Pass19) and applied it to the first 50 stars listed in their Table~B.1. 
The stellar sample that we used in this work is presented in Section~\ref{Results} together with the results. 

The spectra were observed with the CARMENES\footnote{Calar Alto high-Resolution search for M dwarfs with Exo-earths with Near-infrared and optical \'Echelle Spectrographs, \url{http://carmenes.caha.es}} instrument, installed on the Zeiss 3.5\,m telescope at the Calar Alto Observatory, Spain. 
CARMENES combines two highly stable fiber-fed spectrographs covering a spectral range from 520 to 960\,nm in the visible (VIS) and from 960 to 1710\,nm in the near-infrared (NIR), with spectral resolutions of $R \approx$ 94\,600 and 80\,500, respectively \citep[][]{Quirrenbach2018,Reiners2018a}.
The primary goal of this instrument is to search for Earth-sized planets in the habitable zones of M dwarfs. 

For a detailed description on data reduction we refer to \cite{Zechmeister2014}, \cite{Caballero2016}, and Pass19. 
As in the latter we used the high-S/N template spectrum for each star. These templates are a byproduct of the CARMENES radial velocity pipeline {\tt serval} 
\citep[SpEctrum Radial Velocity AnaLyser;][]{Zechmeister2018}. In the standard data flow, the code constructs a template for every target star from at least five individual spectra to derive the radial velocities of a single spectrum by least-square fitting against the template. For our sample, this results in an average S/N of around 159 for the VIS and 328 for the NIR. 

Before creating the templates, the near-infrared spectra were corrected for telluric lines. We did not use the telluric correction for the visible spectra because the telluric features are negligible in the investigated ranges. The telluric correction is explained in detail by \cite{Nagel2020}. A telluric absorption spectrum was modeled for each observation using the telluric-correction tool {\tt Molecfit} \citep{Kausch2014,Smette2015} and then subtracted from the observed spectrum. The result is a telluric-free observed spectrum that can then be used to construct a template.

\subsection{Synthetic model grid}
\label{Analysis.grid}

To train the neural network, we used synthetic model spectra.
The advantage of this approach is that we could generate a large enough number of model spectra and did not have to rely on a limited sample of observations with well known stellar parameters. 
On the other hand, although significant improvements were made in the past years \citep{Allard2013,Husser2013}, synthetic models still cannot fully model stellar atmospheres, especially in the low-temperature range. This is shown by Pass19, for example. 

In this work we used the {\tt PHOENIX} atmosphere code \citep{Hauschildt1992,Hauschildt1993,HauschildtBaron1999}, in particular the PHOENIX-ACES models presented by \cite{Husser2013}. 
The code generates one-dimensional model atmospheres.
They can be computed in local thermodynamical equilibrium (LTE) or non-LTE radiative transfer mode for main sequence stars, brown and white dwarfs, giants, accretion disks, and expanding envelopes of novae and supernovae. As an end product, one- or three-dimensional synthetic spectra can be calculated. 

Several model atmosphere grids used the PHOENIX code as a basis. This includes the NextGen  \citep{Hauschildt1999}, AMES \citep{Allard2001}, BT-Settl \citep{Allard2011}, and PHOENIX-ACES models \citep{Husser2013}. The latter ones were especially designed for cool dwarfs ($T_{\rm eff} \geq 2300$\,K), as they used a new equation of state that accounts for molecule formation in low-temperature stellar atmospheres. The grid that we used in this work is based on the PHOENIX-ACES model grid. 

\begin{table}[!ht]
\caption{PHOENIX grid for DL training.}
\label{tab:grid}
\centering %
    \begin{tabular}{l ccc}
        \hline 
        \hline 
        \noalign{\smallskip}
        Parameter & Minimum & Maximum & Step size\\
        \noalign{\smallskip}
  	    \hline 
        \noalign{\smallskip}
        $T_{\rm eff}$ [K] & 2300 & 4500 & 25 \\
        $\log{g}$ [dex] & 4.2 & 5.5 & 0.1  \\
        {[M/H]} [dex] & --1.0 & +0.8 & 0.1 \\
        $v \sin{i}$ [km\,s$^{-1}$] & 1.5 & 3.0 & 0.5 \\ 
         & 3.0 & 6.0 & 1.0 \\
         & 6.0 & 10.0 & 2.0 \\
         & 10.0 & 60.0 & 5.0 \\
    \noalign{\smallskip}
    \hline
    \end{tabular}
\end{table}

The existing grid (step size of 100\,K for $T_{\rm eff}$, and 0.5\,dex for $\log{g}$ and [M/H]) of stellar spectra was linearly interpolated between the grid points using {\tt pyterpol} \citep{Nemravov2016}. As shown in Pass19, linear interpolation between these grid points produces synthetic spectra that are numerically equivalent to results of simulated spectra. The final grid characteristics used for analyzing the DL modeling capabilities can be seen in Table~\ref{tab:grid}. However, to train models to be applied to the observed spectra, additional restrictions must be considered because the whole grid can provide extra combinations of parameters that are not realistic for M stars. The restrictions regarding the relationship between $T_{\rm eff}$, $\log{g}$, [M/H], and age of the stars were implemented according to~\cite{Bressan2012} and are explained in Section~\ref{Results.CARM}. 

\begin{table}[!ht]
  \caption{Analyzed spectral windows. }
  \label{tab:DAT_WIN}
  \centering %
  \begin{tabular}{ccc}
    \hline 
    \hline 
        \noalign{\smallskip}
     $\lambda_{\rm start}$ [\AA] & $\lambda_{\rm end}$ [\AA] & Chunk size [$\lambda$ points]\\
        \noalign{\smallskip}
	   \hline 
        \noalign{\smallskip}
	   7050 & 7075 & 2048 \\ 
	   7081 & 7115 & 2048 \\ 
	   7121 & 7175 & 4096 \\ 
	   7640 & 7725 & 4096 \\ 
	   8160 & 8225 & 4096 \\ 
	   8401 & 8480 & 4096 \\ 
	   8485 & 8530 & 4096 \\
	   8640 & 8710 & 4096  \\
	   8800 & 8835 & 2048 \\
	        &      & 1024 \\
	        &      & 512 \\
	        &      & 256 \\
       9719 & 9735 & 1024 \\
       9822 & 9847 & 2048 \\
       10570 & 10600 & 1024 \\
       10650 & 10675 & 1024 \\ 
       10760 & 10790 & 1024 \\ 
       11120 & 11135 & 512 \\ 
       11763 & 11795 & 1024  \\ 
       12510 & 12540 & 2001 \\
             &       & 1024 \\
             &       & 512 \\
             &       & 256 \\
	   15146 & 15175 & 512 \\
        \noalign{\smallskip}
 	  \hline
  \end{tabular}
\end{table}

\subsection{Spectrum preparation}

Before we started the training, the synthetic spectra were adjusted to the observations. We did this in several steps.
First, we accounted for instrumental broadening by convolving the synthetic models with a Voigt profile. Our code is based on {\tt libcerf} \citep{libcerf}. The corresponding values for the Gauss and Lorentz part of the Voigt function were taken from \cite{Nagel2020}, who investigated the instrumental profiles of the CARMENES VIS and NIR channels separately. 

Second, to account for different stellar rotation rates ($v \sin{i}$), the synthetic spectra of the finer grid were broadened using our own broadening function in order to speed up the process. It is a Fortran translation of the {\tt rotational\_convolution} function of {\tt Eniric} \citep{Figueira2016}. We kept a limb darkening coefficient of 0.6, which was the default value proposed in the paper.

Because of the high-S/N of the observed CARMENES spectra (S/N > 150), we decided not to include any noise in the synthetic spectra. By using regression models to derive stellar parameters for hotter stars, \cite{GonzalezMarcos2017} show that adding noise to a spectral training set does not improve the results for S/N > 50.

Since the CARMENES {\tt caracal} pipeline produces only flat-relative normalized spectra, we applied a continuum normalization. We used the Gaussian Inflection Spline Interpolation Continuum ({\tt GISIC}) routine\footnote{\url{https://pypi.org/project/GISIC/}} developed by D.\,D.~Whitten and designed for spectra with strong molecular features. The spectrum was smoothed using a Gaussian, then molecular bands were identified from a numerical gradient and continuum points were selected. A cubic spline interpolation was performed to normalize the continuum over the whole spectral range (see Table~\ref{tab:DAT_WIN}). The same procedure was also applied to the observed CARMENES spectra for each spectral range of interest. To prevent the possible edge effects of the normalization we extended every window by 5\,{\AA} on each side. 

Finally, because of the spatial motion of the stars, an absolute radial-velocity correction wass required for the observed spectra. We used a method similar to the one implemented in {\tt serval}, 
which employs the cross-correlation \citep[{\tt crosscorrRV} from {\tt PyAstronomy},][]{pya} between a PHOENIX model spectrum and the observed spectrum.
By applying the radial velocity correction, the wavelengths were shifted and, therefore, the wavelength grid was different for each CARMENES spectrum. Because we needed a universal wavelength grid in order to apply the DL models, we linearly interpolated the wavelength grid of the observations to the original wavelength grid of the synthetic models. The process ended up with 449\,806 synthetic model spectra for training the DL models.
 

\subsection{Implementation}

Regarding the implementation of the algorithm, we used {\tt TensorFlow v2}, which is the premier open-source DL framework commonly available. It was developed and maintained by Google \citep{Abadi2015} and, since its direct usage can be challenging, a front-end framework named {\tt Keras} was used as well \citep{Chollet2015}.

The adoption of the {\tt TensorFlow} framework for DL model creation enables the usage of accelerated hardware based on the Nvidia general-purpose graphics processing unit (GPU) cards, which outperforms the central processing unit computation (CPU) time by around a factor of twenty \citep{Mittal2019}.
In this application, we used GPU cards with 11\,GB of RAM and 4352 computing cores.
The training time for a model experiencing proper convergence depended on the training data size, but also on the architecture and number of epochs, and it varied between 45 minutes to several hours.

\subsection{Different DL approaches}
\label{Analysis.architecture}


Different DL architectures were considered, some of them inspired by literature such as \cite{Sharma2019}, {\tt StarNet} \citep{kielty2018starnet}, and many other homemade architectures.
To this end, a flexible python code was implemented where the topology for the convolutional structure and for the ANN layers were passed as parameters. In this way it was possible to distribute the computations among different computing nodes to increase parallelism, as well as keeping the software easy to maintain.

\begin{figure*}[!ht]
  \centering
  \includegraphics[width=0.95\textwidth]{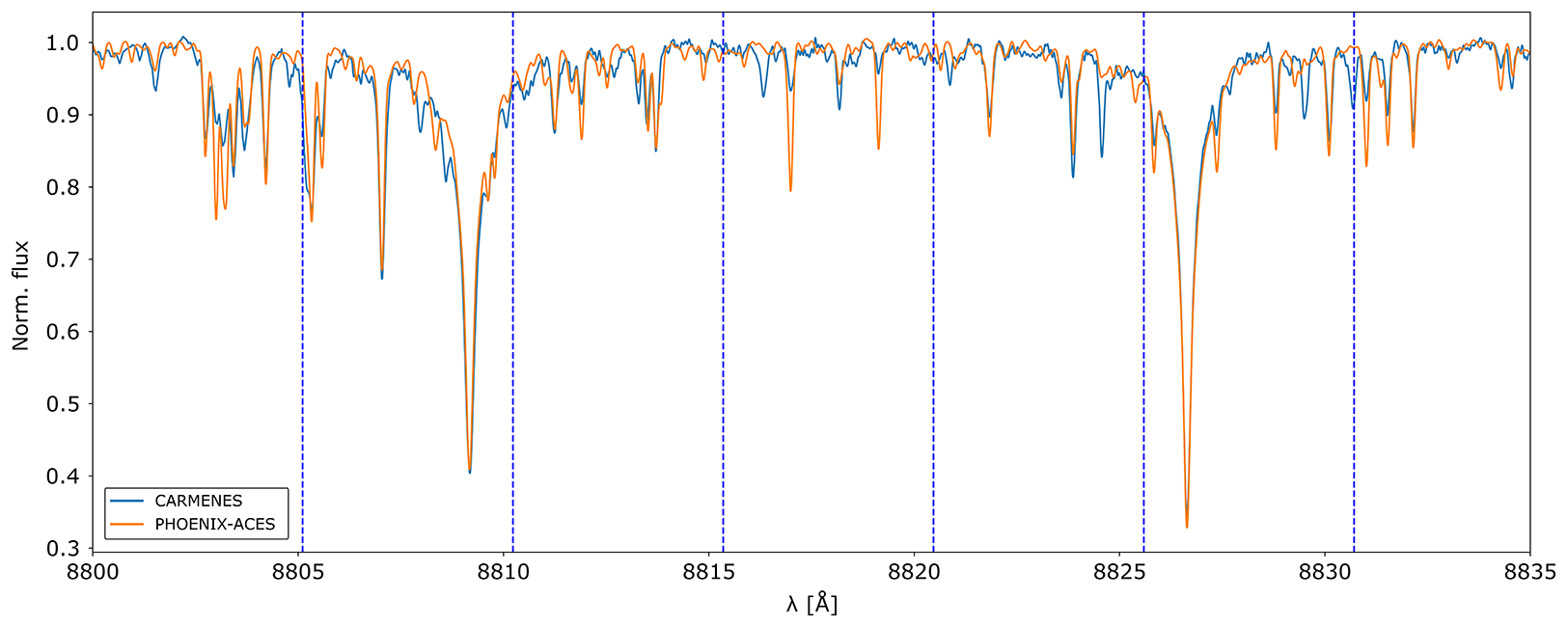}  
  \caption{Example of one spectral window used with different chunks of size 512 points indicated by the vertical dashed lines.}
  \label{fig:window_chunks}
\end{figure*}

We used different spectral windows to derive the models. These were taken from Pass19 and are summarized in Table~\ref{tab:DAT_WIN}. For each window we considered a certain chunk size, which refers to the number of wavelengths points within the window. For some windows we used several chunk sizes, ranging from 256 to 4096 wavelength points, to explore the impact on the DL model and its predictions of the stellar parameters. Figure~\ref{fig:window_chunks} shows an example for such a spectral window with different chunks marked. Three approaches were defined for the analysis as follows.

In approach A, we trained a DL model for one spectral window and predicted each stellar parameter individually. We modified the DL architecture by varying the number of convolutional, pooling, and ANN hidden layers. The different architectures ranged from just three convolutional layers with one max pooling layer, passing the flattened vector of features to an ANN having three hidden layers, to eight convolutional layers with and without intermediate pooling layers, both maximum and averaged. The feature vector fed into an ANN having five hidden layers. The convolutional layers had implemented strategies from a few replications per layer and increasing as they moved over the layers and vice versa. 
In this way we investigated a possible difference in predictions depending on the architecture and spectral window used. 

In approach B, different combinations of stellar parameters were analyzed. In this approach we derived a DL model for individual parameters, for both $T_{\rm eff}$ and $\log{g}$, and for [M/H] and $v \sin{i}$, as well as for all the four parameters at the same time. 
Also here we determined predictions for the spectral windows in Table~\ref{tab:DAT_WIN} and investigated different architectures. From this approach we see how a combined DL model impacted the stellar parameter predictions.

In approach C, we derived a DL model for each stellar parameter separately, but we combined spectral windows. For example, we combined the window starting at 8800~{\AA} with the window starting at 12510~{\AA} to investigate the differences in the predictions when more spectral information is used. Also, other window couplings were analyzed, such as the combination of all the VIS channel windows, and all the NIR channel windows.
Again, we determined DL models for different architectures.

\begin{figure*}[!ht]
  \centering
  \includegraphics[width=0.48\linewidth]{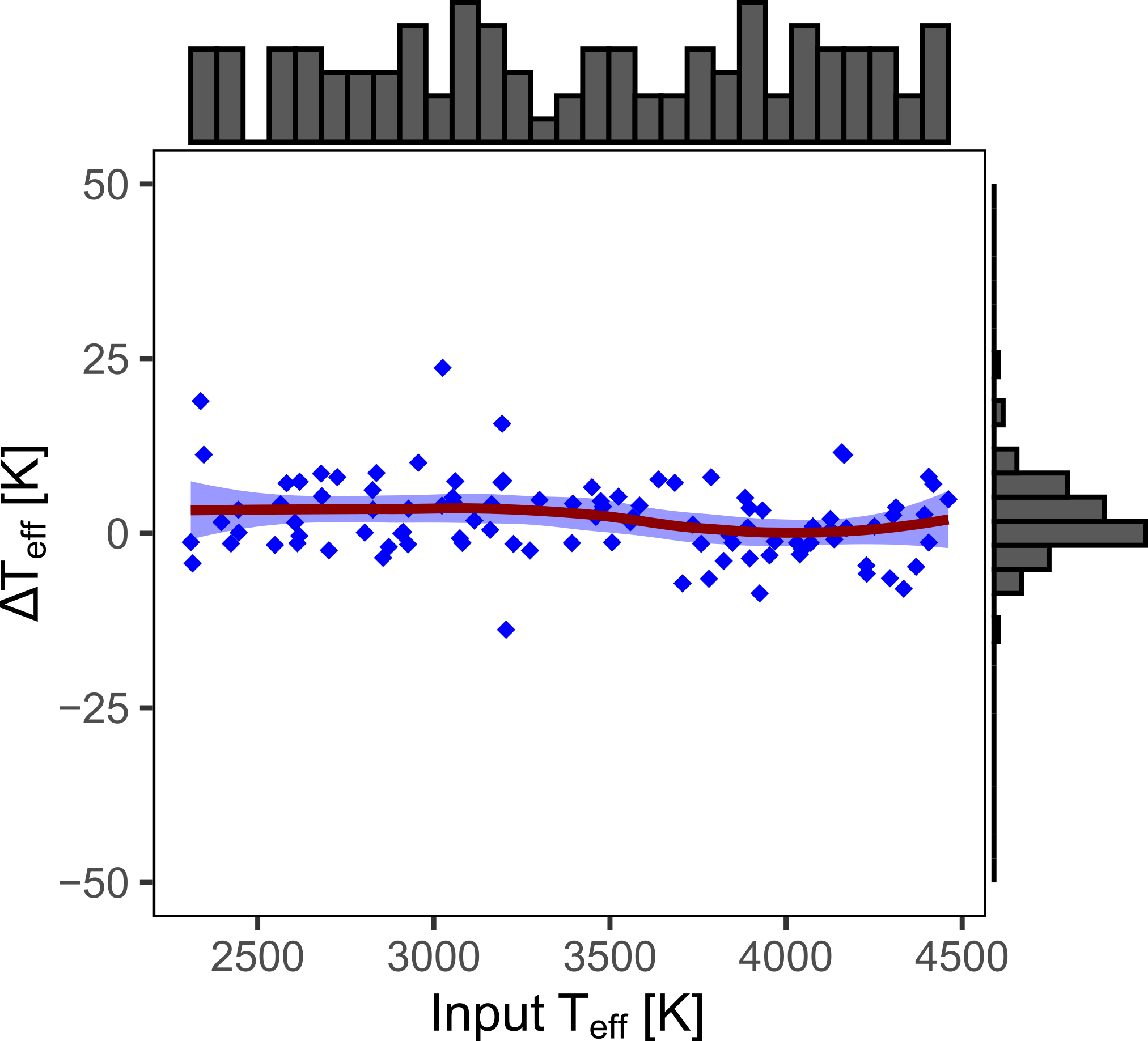}\hfil
  \includegraphics[width=0.48\linewidth]{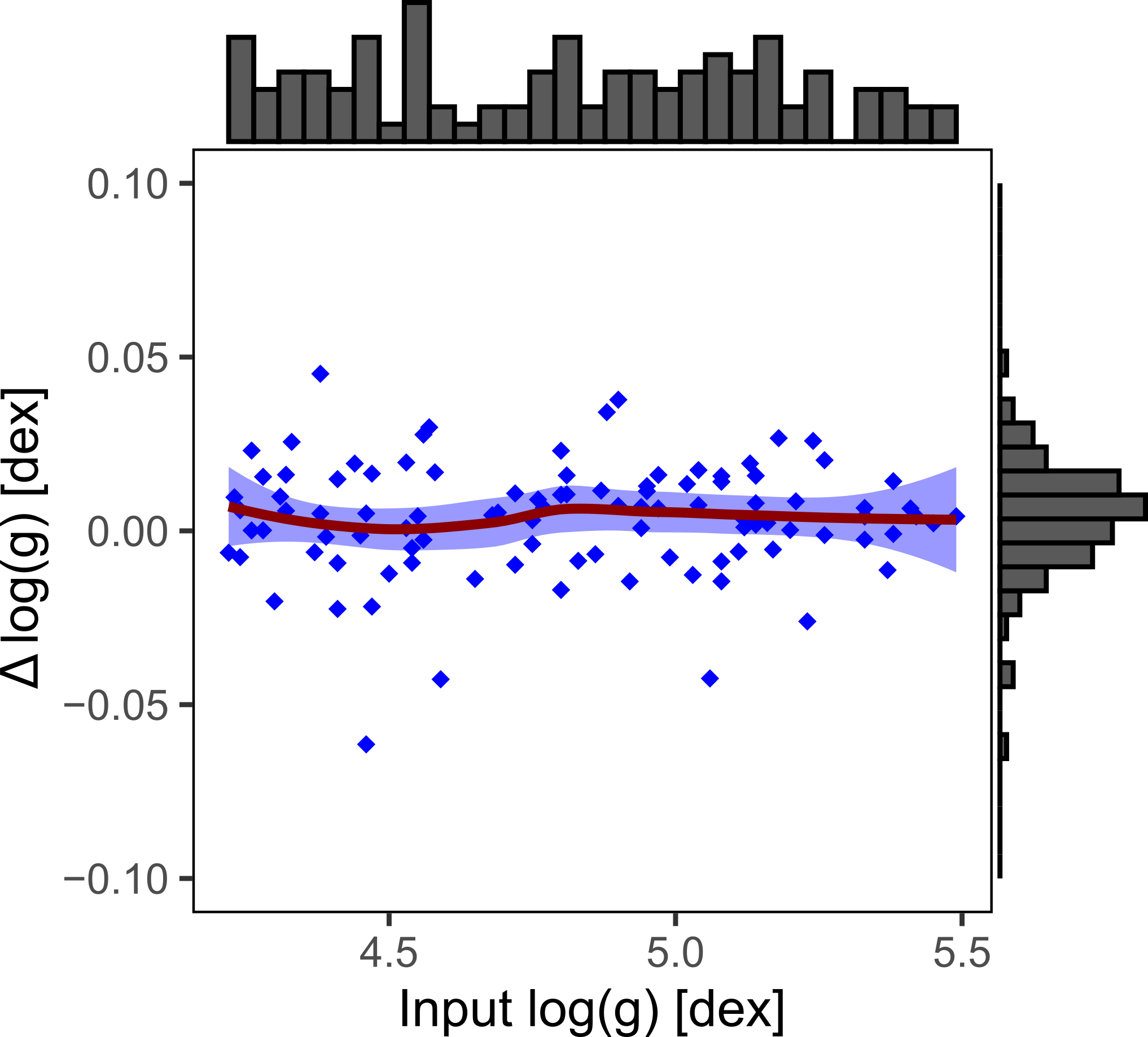}\par\medskip
  \includegraphics[width=0.48\linewidth]{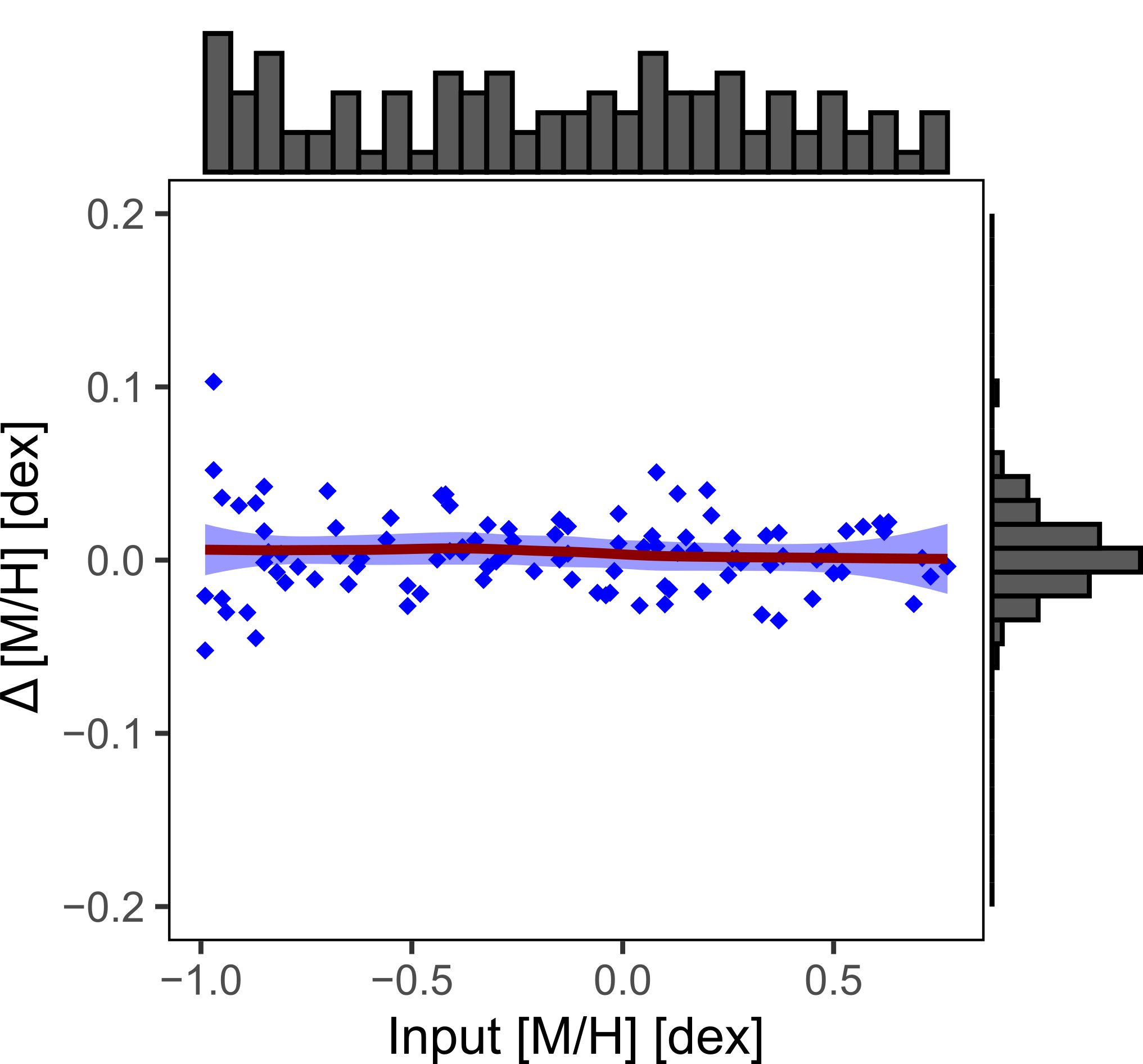}\hfil
  \includegraphics[width=0.48\linewidth]{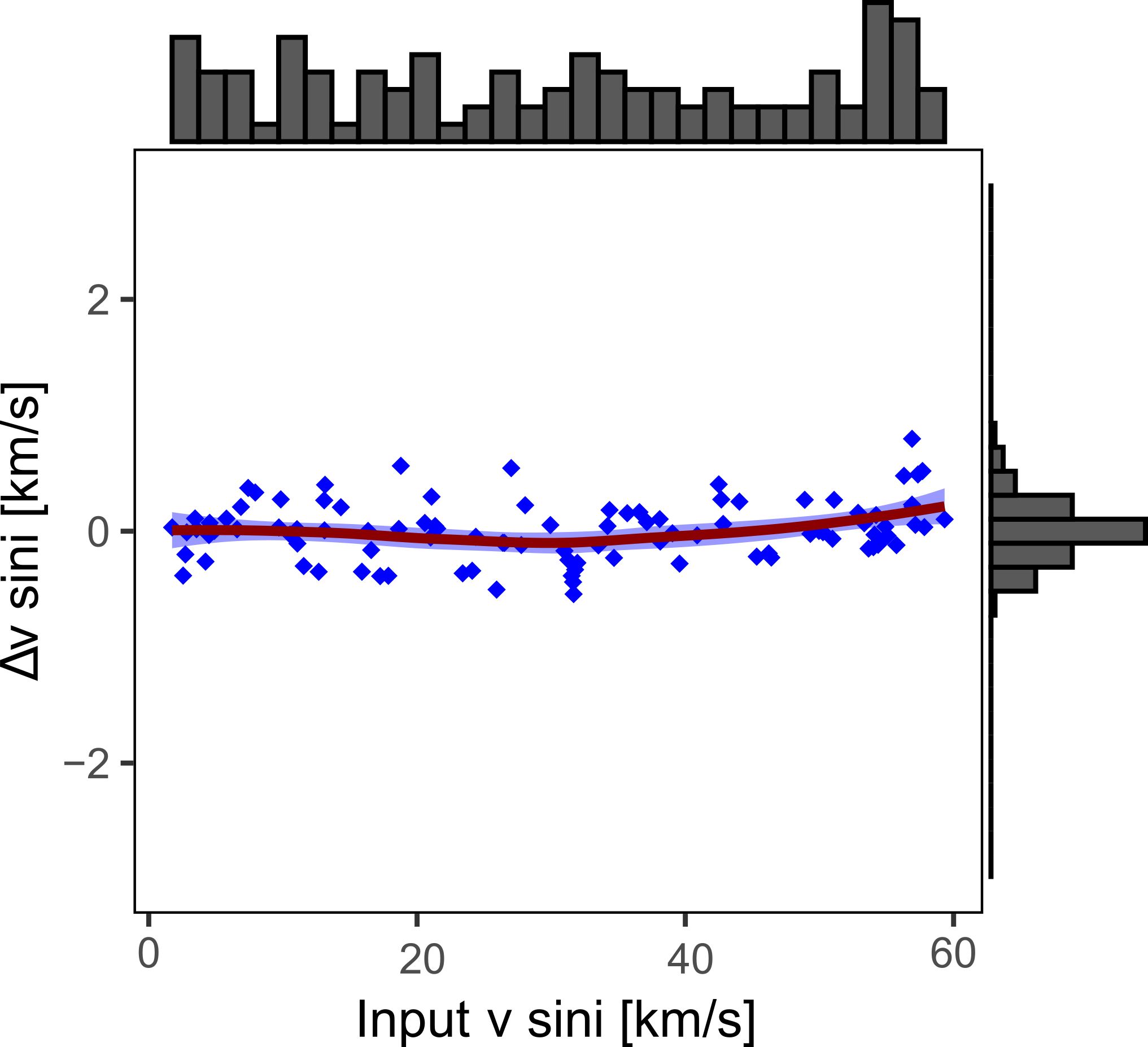}
\caption{Differences between input and output stellar parameters for the test set for an example DL model. The red line shows the average of all points and the blue shaded area is the 95\,\% confidence region.}
\label{fig:DLTest}
\end{figure*}

As a summary, more than 2000 different DL models were created, and for selecting the most suitable ones, specific quality criteria were defined. Every model was applied across the test sample to estimate the quality of the forecast. This error threshold was used as quality criteria to select them for further usage over the observed spectra. The number of selected models, as well as the quality criteria, are shown in Table~\ref{tab:Threshold_test}. The thresholds for defining a good quality model were adopted from the MSE criterion, ranging between $5 \cdot 10^{-4}$ and $10^{-5}$. The aim was to produce a difference between real and estimated parameters less than the threshold presented in Table~\ref{tab:Threshold_test}, independently of the value of the parameter. As an example, the prediction quality for the test dataset for some of the models is plotted in Fig.~\ref{fig:DLTest}.

\begin{table}
  \caption{Quality criteria and number of suitable models trained from PHOENIX.}
  \label{tab:Threshold_test}
  \centering %
  \begin{tabular}{lcccc}
    \hline 
    \hline 
        \noalign{\smallskip}
     Parameter & Error & N\textsuperscript{o} & N\textsuperscript{o} & N\textsuperscript{o} \\
     & threshold & models & architectures & windows \\
        \noalign{\smallskip}
	   \hline 
        \noalign{\smallskip}
	   $T_{\rm eff}$ & 25\,K & 89 & 22 & 11 \\
	   $\log{g}$ & 0.05\,dex & 101 & 15 &  7 \\
	   {[M/H]} & 0.05\,dex & 95 & 18 & 9 \\
	   $v \sin{i}$  & 1\,km\,s$^{-1}$ & 113 & 16 & 10 \\
        \noalign{\smallskip}
 	  \hline
  \end{tabular}
\end{table}


\subsection{Application to CARMENES spectra}
\label{Analysis.DL+CARMENES}

The synthetic gap is well-known in ML and refers to the differences in feature distributions between synthetic and observed data \citep{Fabbro2018}. Because the spectra are high-dimensional data with, in our case of synthetic spectra, up to 4096 flux values (i.e., dimensions) per spectral window, dimensional reduction is necessary to visualize the data in low-dimensional space. So we verified the synthetic gap by using a nonlinear projector from high-dimensional flux space into a two-dimensional space that preserved the local topology in order to understand similarities between PHOENIX and CARMENES spectra families. We decided to use the Uniform Manifold Approximation and Projection \citep[UMAP;][]{mcinnes2018umap-software} with a metric that is a correlation between the spectra.

\begin{figure}[!ht]
  \centering
  \includegraphics[width=0.49\textwidth]{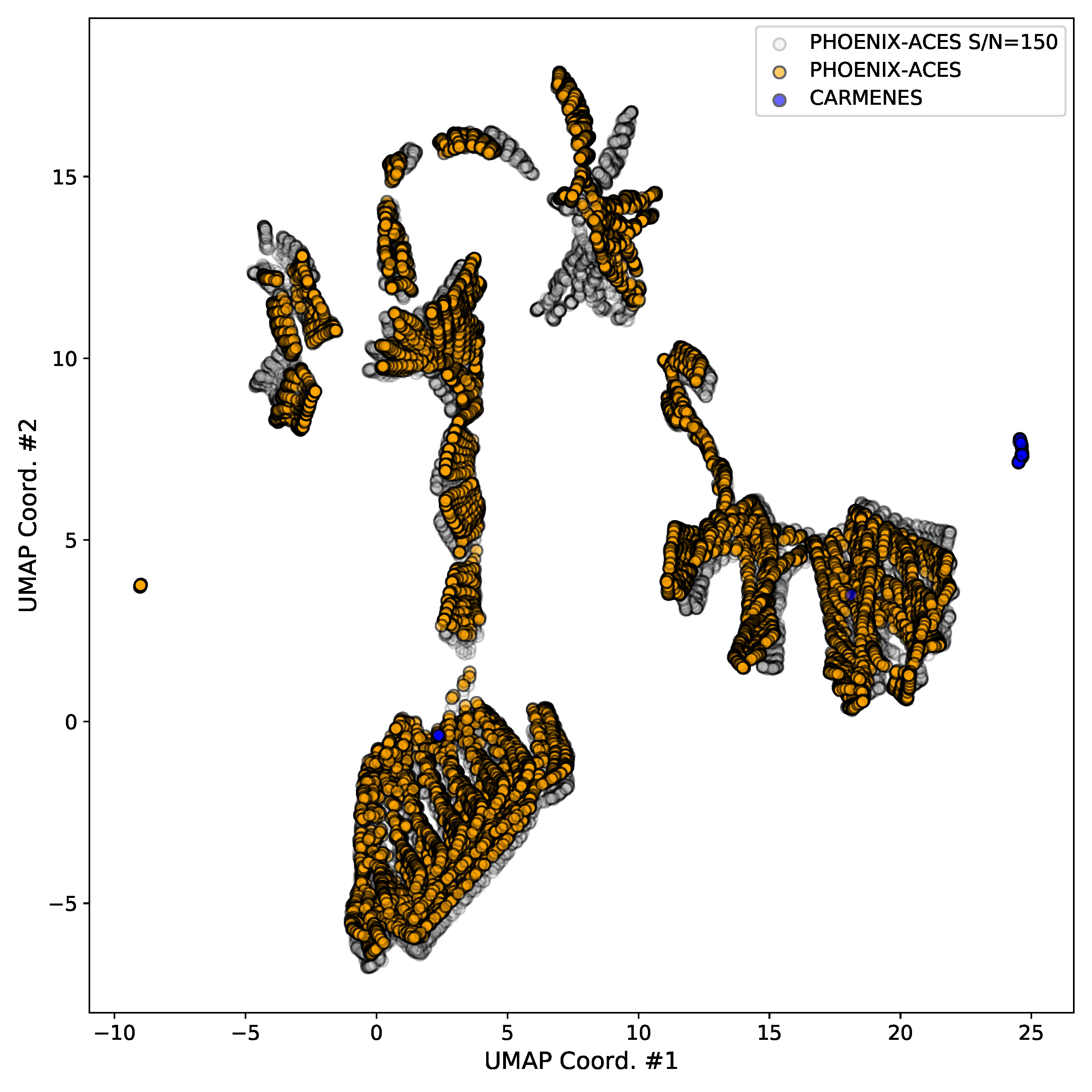}  
  \caption{Representative UMAP two-dimensional projection of CARMENES (blue) and PHOENIX spectra (S/N = ${\infty}$ in yellow, S/N = 150 in gray) built with the flux values from the 8800--8835\,{\AA} window. Only four CARMENES spectra show similarities with the PHOENIX feature distribution, while the rest are considerably different. Other spectral windows show a similar behavior.}
  \label{fig:UMAP}
\end{figure}

As can be seen in Fig.~\ref{fig:UMAP}, only four of our investigated CARMENES spectra are close to the PHOENIX family (we plotted only a subset of the full PHOENIX grid for visibility). However, there is a significant set of CARMENES spectra (46 of 50) far away from the PHOENIX sample set used to train the DL, meaning that the flux features described by synthetic and observed spectra are significantly different. These large deviations can be explained by the synthetic gap. Due to the fact that synthetic spectra are not perfect, it is expected that the feature distributions of synthetic and observed spectra do not match completely, meaning that not all CARMENES spectra would fall right within the PHOENIX spectra in Fig.~\ref{fig:UMAP}. However, such a big difference as observed here indicates significant differences between those two spectral families and will transform into higher uncertainties on the stellar parameters derived using these synthetic spectra.
To assess the possible relevance of the noise in this work we also added white Gaussian noise (S/N = 50, 100, 150) to the PHOENIX set and compared the resulting UMAP projections. All the noisy PHOENIX spectra show a similar behavior as the noise-free spectra. In our case, an S/N of 150 served as a lower limit for the CARMENES spectra, since most of the observed spectra used in this work have higher S/N. 
The noisy PHOENIX spectra are not closer to the observed data in the UMAP, indicating that noise is not responsible for the synthetic gap.
We note that, since this is a dimensional projection, the axes in Fig.~\ref{fig:UMAP} are not labeled, because they do not have a specific meaning here.

\begin{figure*}[!ht]
  \centering
  \includegraphics[width=0.9\textwidth]{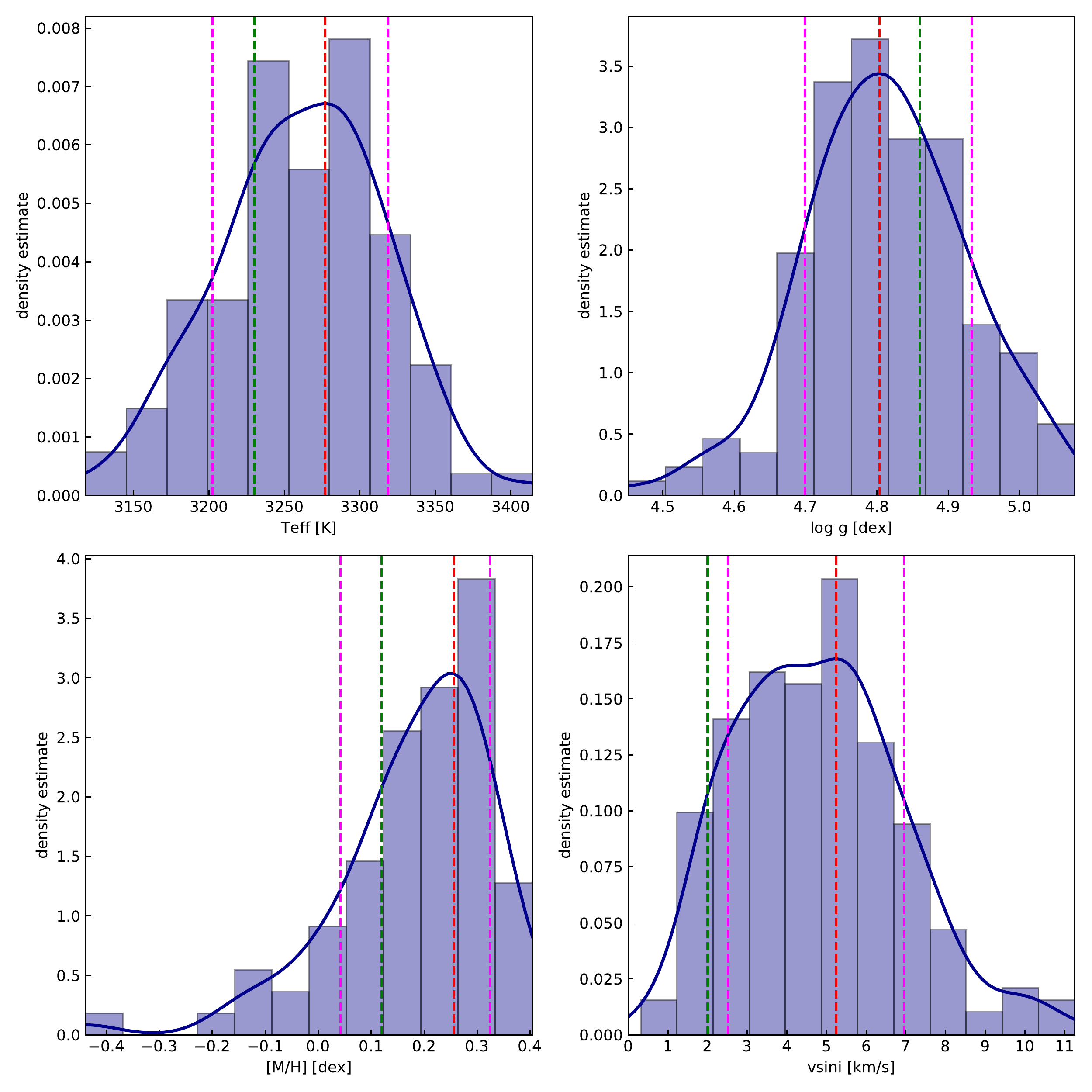}  
  \caption{DL estimations distribution for the CARMENES star GJ~169.1A (J04311+589). The KDE maximum (i.e., our adopted estimations, red lines) is shown together with the 1$\sigma$ uncertainties (magenta lines) and results from Pass19 (green lines). The dark blue curve represents the Gaussian kernel density estimate. We used the default rules from the {\tt seaborn distplot} function: Freedman-Diaconis' \citep{FreedmanDiaconis1981} for the histogram bin width and Scott's rule \citep{Scott1979} for the kernel size.}
  \label{fig:KDEestimations}
\end{figure*}

Being aware of such a gap, all accepted DL models from all three DL approaches described in Sect.~\ref{Analysis.architecture} were applied to the 50 CARMENES spectra in order to estimate the stellar parameters. 
The parameter estimations from each DL model were collected and the probability density function was determined using the Kernel Density Estimate \citep[KDE;][]{Rosenblatt1956, Parzen1962}.
Based on such a density of probability function, the maximum was retained as the confident estimation for the parameter. This was done for each star and each stellar parameter separately. For providing the uncertainty for each star and parameter, the 1$\sigma$ thresholds of the predictions were calculated. An example for a representative star is presented in Fig.~\ref{fig:KDEestimations}. We included the stellar parameters derived by Pass19 for comparison.

\section{Results and discussion}
\label{Results}

\subsection{Performance of different DL approaches}
\label{Results.Architecture}
When we apply our three DL approaches (A, B, C) to the PHOENIX test set, the predicted stellar parameters show that there is only little, if any, influence from the DL architecture, as all of them are able to produce good quality DL models.
There are also minimal differences when considering stellar parameters derived from different windows. Therefore, we conclude that there is enough spectral information to determine stellar parameters from any window, no matter what chunk size, since also no particular improvement is found when several windows were joined. 
The DL models are able to successfully predict individual stellar parameters.
Obviously, this only refers to PHOENIX models and cannot be translated to the observed spectra because of the particular effects of the synthetic gap, and some other factors, such as measuring at specific wavelength ranges, telluric correction or different S/N.
Nevertheless, when estimating the information content of stellar spectra from a purely theoretical point of view, similar results are obtained. For example, \citet{Hafner1994} estimated to be able to retrieve more than ten parameters in a 1000\,{\AA} wide chunk in which the resolution is only a quarter of that of CARMENES.
Therefore, a key lesson learned is related to the capability of the DL models to disentangle the specific effects found in the spectra by individual parameters even if a very small chunk of the spectrum is considered. Indeed, no particular attention to the DL architecture is required as all of those tested are capable of performing well. 

During our tests, we also trained models predicting all four stellar parameters simultaneously, as it is done by {\tt The Cannon}. However, we find that these models always give worse predictions, meaning higher validation errors, than individual ones. Therefore, we decided to estimate each stellar parameter individually.

\subsection{Performance on CARMENES spectra}
\label{Results.CARM}

To address the second goal of this work (evaluate its application to observed spectra and the impact of the synthetic gap on the estimation of stellar parameters), we applied all trained models matching the quality criteria (see Table~\ref{tab:Threshold_test}) to the observed CARMENES spectra, collected the results for each star, and drew statistical distributions for each parameter. This technique has the advantage of estimating uncertainties for each star and stellar parameter by calculating the 1$\sigma$ deviations for each distribution. 

\begin{figure}[!ht]
  \centering
  \includegraphics[width=0.49\textwidth]{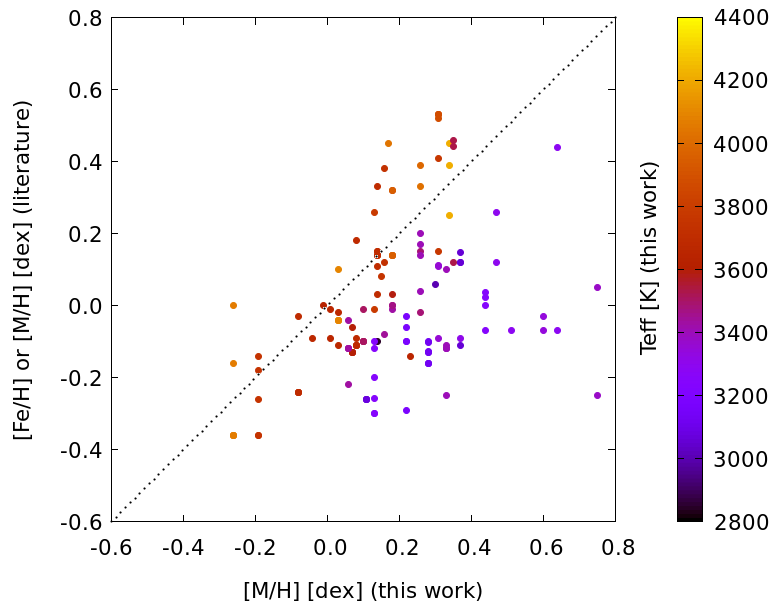}\hfil
   \includegraphics[width=0.49\textwidth]{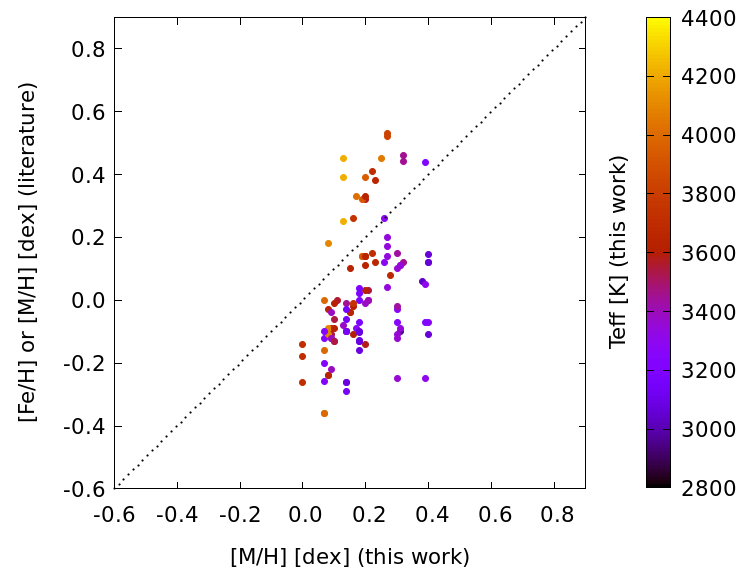}\par\medskip
\caption{Literature comparison for [M/H] with $T_{\rm eff}$ color-coded, before ({\it top}) and after ({\it bottom}) applying the PARSEC constraint on the synthetic grid.}
\label{fig:degeneracy}
\end{figure}

Figure~\ref{fig:degeneracy} shows a comparison between our estimated [M/H] and literature values, color-coded with our derived $T_{\rm eff}$. The degeneracy at low $T_{\rm eff}$ and high [M/H] is evident in the top panel. 
This behavior was described already, for example in \cite{Passegger2018} and Pass19, who decided to break the degeneracy by determining $\log{g}$ independently from evolutionary models. In contrast to Pass19, who used the PARSEC v1.2S evolutionary 
models \citep{Bressan2012,Chen2014,Chen2015,Tang2014} to derive $\log{g}$ depending on $T_{\rm eff}$ and metallicity suggested by their downhill simplex method, we used the PARSEC library to constrain our synthetic model grid with which we trained our DL models. 
In this way we could remove stellar parameter combinations that are physically unrealistic for M dwarfs (i.e., they correspond to objects far away from the main sequence) and helped the DL to break the degeneracy. 
After applying this constraint, we tested our improved approach and trained new DL models for the wavelength window 8800--8835\,{\AA}, since this window shows the smallest MSE from all windows we investigated. Using the same quality criteria as presented in Table~\ref{tab:Threshold_test}, we ended up with more than 200 accepted DL models for each stellar parameter. We find that constraining the synthetic model grid is indeed capable of breaking the observed parameter degeneracy (see bottom panel of Fig.~\ref{fig:degeneracy}). The hereby estimated stellar parameters, together with their uncertainties for the 50 CARMENES stars, are presented in Table~\ref{tab:results}.


\begin{table*}
\caption[]{Sample of CARMENES stars and derived parameters with 1 $\sigma$ uncertainties.}
\label{tab:results}
\centering
    \begin{tabular}{lllccccc}
        \hline 
        \hline 
        \noalign{\smallskip}
       Karmn  & Name & GJ &  Spectral type & $T_{\rm eff}$ [K] & $\log{g}$ [dex] & [M/H] [dex] & $v\sin{i}$ [kms$^{-1}$]\\
        \noalign{\smallskip}
  	    \hline 
        \noalign{\smallskip}
J00051+457	&	 GJ 2 	&	 2 	&	 M1.0\,V 	&	3738$_{-35}^{+60}$	&	4.68$_{-0.07}^{+0.03}$	&	0.16$_{-0.12}^{+0.06}$	&	4.1$_{-1.0}^{+1.7}$	\\
J00067--075	&	 GJ 1002  	&	 1002 	&	 M5.5\,V 	&	2949$_{-87}^{+114}$	&	5.13$_{-0.22}^{+0.06}$	&	0.31$_{-0.44}^{+0.16}$	&	4.9$_{-3.0}^{+3.0}$	\\
J00162+198E	&	 LP 404-062  	&	 1006B 	&	 M4.0\,V 	&	3340$_{-84}^{+62}$	&	4.80$_{-0.09}^{+0.15}$	&	0.31$_{-0.26}^{+0.09}$	&	4.3$_{-1.7}^{+2.6}$	\\
J00183+440	&	 GX And 	&	 15A 	&	 M1.0\,V 	&	3704$_{-43}^{+57}$	&	4.71$_{-0.06}^{+0.05}$	&	0.00$_{-0.08}^{+0.08}$	&	3.2$_{-0.8}^{+1.8}$	\\
J00184+440	&	 GQ And 	&	 15B 	&	 M3.5\,V 	&	3223$_{-46}^{+69}$	&	4.96$_{-0.09}^{+0.12}$	&	0.07$_{-0.20}^{+0.11}$	&	5.7$_{-2.8}^{+1.3}$	\\
J00286--066	&	 GJ 1012 	&	 1012 	&	 M4.0\,V 	&	3430$_{-54}^{+81}$	&	4.74$_{-0.08}^{+0.12}$	&	0.30$_{-0.20}^{+0.07}$	&	4.5$_{-1.8}^{+2.0}$	\\
J00389+306	&	 Wolf 1056 	&	 26 	&	 M2.5\,V 	&	3592$_{-43}^{+57}$	&	4.72$_{-0.07}^{+0.07}$	&	0.16$_{-0.12}^{+0.05}$	&	4.1$_{-1.5}^{+1.3}$	\\
J00570+450	&	 G 172-030 	&	 ... 	&	 M3.0\,V 	&	3372$_{-46}^{+57}$	&	4.82$_{-0.06}^{+0.13}$	&	0.20$_{-0.16}^{+0.07}$	&	4.5$_{-1.7}^{+1.6}$	\\
J01013+613	&	 GJ 47 	&	 47 	&	 M2.0\,V  	&	3532$_{-30}^{+52}$	&	4.77$_{-0.07}^{+0.09}$	&	0.09$_{-0.13}^{+0.06}$	&	4.1$_{-1.4}^{+1.3}$	\\
J01025+716	&	 BD+70 68 	&	 48 	&	 M3.0\,V 	&	3668$_{-53}^{+76}$	&	4.66$_{-0.06}^{+0.07}$	&	0.28$_{-0.14}^{+0.06}$	&	4.1$_{-1.3}^{+1.8}$	\\
J01026+623	&	 BD+61 195 	&	 49 	&	 M1.5\,V 	&	3698$_{-34}^{+58}$	&	4.64$_{-0.02}^{+0.08}$	&	0.22$_{-0.12}^{+0.07}$	&	4.4$_{-1.0}^{+1.8}$	\\
J01048--181	&	 GJ 1028  	&	 1028 	&	 M5.0\,V 	&	3027$_{-73}^{+96}$	&	5.04$_{-0.19}^{+0.09}$	&	0.38$_{-0.40}^{+0.12}$	&	4.5$_{-2.6}^{+3.3}$	\\
J01125--169	&	 YZ Cet 	&	 54.1 	&	  M4.5\,V 	&	3075$_{-57}^{+68}$	&	5.06$_{-0.10}^{+0.11}$	&	0.14$_{-0.31}^{+0.12}$	&	6.0$_{-2.3}^{+1.9}$	\\
J01339--176	&	 LP 768-113 	&	 ... 	&	 M4.0\,V 	&	3268$_{-59}^{+37}$	&	4.94$_{-0.07}^{+0.11}$	&	0.17$_{-0.21}^{+0.07}$	&	5.4$_{-1.2}^{+2.0}$	\\
J01433+043	&	 GJ 70 	&	 70 	&	 M2.0\,V 	&	3531$_{-37}^{+60}$	&	4.75$_{-0.07}^{+0.09}$	&	0.10$_{-0.11}^{+0.06}$	&	4.3$_{-1.5}^{+1.5}$	\\
J01518+644	&	 G 244-037 	&	 3117A 	&	 M2.5\,V 	&	3581$_{-39}^{+51}$	&	4.70$_{-0.05}^{+0.09}$	&	0.20$_{-0.12}^{+0.06}$	&	4.4$_{-1.6}^{+1.3}$	\\
J02002+130	&	 TZ Ari 	&	 83.1 	&	 M3.5\,V 	&	3074$_{-62}^{+60}$	&	5.07$_{-0.10}^{+0.10}$	&	0.18$_{-0.32}^{+0.11}$	&	6.4$_{-2.0}^{+2.2}$	\\
J02015+637	&	 G 244-047 	&	 3126 	&	 M3.0\,V 	&	3528$_{-48}^{+58}$	&	4.72$_{-0.07}^{+0.09}$	&	0.21$_{-0.14}^{+0.06}$	&	4.0$_{-1.5}^{+1.5}$	\\
J02070+496	&	 G 173-037 	&	 ... 	&	 M3.5\,V 	&	3382$_{-56}^{+36}$	&	4.86$_{-0.06}^{+0.13}$	&	0.13$_{-0.16}^{+0.07}$	&	5.2$_{-1.8}^{+1.3}$	\\
J02088+494	&	 G 173-039  	&	 3136 	&	 M3.5\,V 	&	3300$_{-27}^{+23}$	&	4.87$_{-0.08}^{+0.06}$	&	0.39$_{-0.13}^{+0.09}$	&	25.9$_{-1.1}^{+1.9}$	\\
J02123+035	&	 BD+02 348 	&	 87 	&	 M1.5\,V 	&	3990$_{-68}^{+106}$	&	4.61$_{-0.05}^{+0.05}$	&	0.07$_{-0.13}^{+0.08}$	&	3.2$_{-1.2}^{+2.1}$	\\
J02222+478	&	 BD+47 612 	&	 96 	&	 M0.5\,V 	&	3926$_{-51}^{+70}$	&	4.61$_{-0.04}^{+0.05}$	&	0.19$_{-0.16}^{+0.07}$	&	4.4$_{-1.2}^{+1.9}$	\\
J02336+249	&	 GJ 102 	&	 102 	&	 M4.0\,V 	&	3180$_{-65}^{+36}$	&	5.03$_{-0.10}^{+0.08}$	&	0.18$_{-0.24}^{+0.09}$	&	6.9$_{-1.5}^{+2.2}$	\\
J02358+202	&	 BD+19 381 	&	 104 	&	 M2.0\,V 	&	3648$_{-46}^{+55}$	&	4.67$_{-0.05}^{+0.08}$	&	0.23$_{-0.16}^{+0.06}$	&	3.8$_{-1.0}^{+2.0}$	\\
J02362+068	&	 BX Cet 	&	 105B 	&	 M4.0\,V 	&	3354$_{-55}^{+85}$	&	4.80$_{-0.10}^{+0.12}$	&	0.30$_{-0.24}^{+0.07}$	&	4.5$_{-1.6}^{+2.4}$	\\
J02442+255	&	 VX Ari 	&	 109 	&	 M3.0\,V 	&	3442$_{-45}^{+55}$	&	4.79$_{-0.06}^{+0.12}$	&	0.14$_{-0.15}^{+0.05}$	&	4.3$_{-1.6}^{+1.5}$	\\
J02519+224	&	 RBS 365  	&	 ... 	&	 M4.0\,V 	&	3253$_{-33}^{+21}$	&	4.81$_{-0.05}^{+0.06}$	&	0.40$_{-0.11}^{+0.07}$	&	28.8$_{-1.4}^{+1.5}$	\\
J02565+554W	&	 Ross 364 	&	 119A 	&	 M1.0\,V 	&	4048$_{-68}^{+90}$	&	4.59$_{-0.06}^{+0.04}$	&	0.25$_{-0.17}^{+0.10}$	&	4.6$_{-1.7}^{+2.3}$	\\
J03133+047	&	 CD Cet 	&	 1057 	&	 M5.0\,V 	&	3056$_{-68}^{+84}$	&	4.98$_{-0.18}^{+0.10}$	&	0.40$_{-0.38}^{+0.10}$	&	4.8$_{-2.7}^{+3.2}$	\\
J03181+382	&	 HD 275122 	&	 134 	&	 M1.5\,V 	&	3819$_{-51}^{+66}$	&	4.61$_{-0.03}^{+0.06}$	&	0.27$_{-0.15}^{+0.10}$	&	4.3$_{-1.3}^{+2.0}$	\\
J03213+799	&	 GJ 133 	&	 133 	&	 M2.0\,V 	&	3609$_{-36}^{+57}$	&	4.72$_{-0.06}^{+0.07}$	&	0.09$_{-0.11}^{+0.06}$	&	3.8$_{-1.1}^{+1.6}$	\\
J03217--066	&	 G 077-046 	&	 3218 	&	 M2.0\,V 	&	3577$_{-33}^{+53}$	&	4.71$_{-0.05}^{+0.09}$	&	0.20$_{-0.12}^{+0.06}$	&	4.6$_{-1.4}^{+1.3}$	\\
J03463+262	&	 HD 23453 	&	 154 	&	 M0.0\,V 	&	3943$_{-53}^{+71}$	&	4.63$_{-0.05}^{+0.03}$	&	0.20$_{-0.13}^{+0.10}$	&	5.2$_{-1.5}^{+1.7}$	\\
J03473--019	&	 G 080-021 	&	 ... 	&	 M3.0\,V 	&	3432$_{-26}^{+28}$	&	4.88$_{-0.07}^{+0.11}$	&	0.20$_{-0.18}^{+0.09}$	&	9.5$_{-1.4}^{+1.4}$	\\
J03531+625	&	 Ross 567 	&	 ... 	&	 M3.0\,V 	&	3553$_{-40}^{+66}$	&	4.75$_{-0.07}^{+0.09}$	&	0.11$_{-0.14}^{+0.07}$	&	4.1$_{-1.6}^{+1.2}$	\\
J04225+105	&	 LSPM J0422+1031 	&	 ... 	&	 M3.5\,V 	&	3445$_{-48}^{+67}$	&	4.70$_{-0.08}^{+0.09}$	&	0.32$_{-0.17}^{+0.06}$	&	4.2$_{-1.5}^{+2.1}$	\\
J04290+219	&	 BD+21 652 	&	 169 	&	 M0.5\,V 	&	4210$_{-89}^{+68}$	&	4.60$_{-0.07}^{+0.03}$	&	0.13$_{-0.18}^{+0.14}$	&	5.9$_{-2.2}^{+2.0}$	\\
J04311+589	&	 STN 2051A 	&	 169.1A 	&	 M4.0\,V 	&	3277$_{-75}^{+42}$	&	4.80$_{-0.10}^{+0.13}$	&	0.26$_{-0.22}^{+0.06}$	&	5.2$_{-2.7}^{+1.7}$	\\
J04376--110	&	 BD-11 916 	&	 173 	&	 M1.5\,V 	&	3633$_{-37}^{+60}$	&	4.71$_{-0.07}^{+0.06}$	&	0.15$_{-0.13}^{+0.06}$	&	3.7$_{-1.3}^{+1.4}$	\\
J04376+528	&	 BD+52 857 	&	 172 	&	 M0.0\,V 	&	4090$_{-65}^{+61}$	&	4.63$_{-0.05}^{+0.02}$	&	0.08$_{-0.14}^{+0.09}$	&	4.9$_{-1.3}^{+1.7}$	\\
J04429+189	&	 HD 285968 	&	 176 	&	 M2.0\,V 	&	3635$_{-44}^{+53}$	&	4.68$_{-0.05}^{+0.07}$	&	0.20$_{-0.12}^{+0.06}$	&	4.1$_{-1.4}^{+1.5}$	\\
J04429+214	&	 2M J04425586+2128230 	&	 ... 	&	 M3.5\,V 	&	3396$_{-43}^{+71}$	&	4.78$_{-0.06}^{+0.13}$	&	0.21$_{-0.15}^{+0.08}$	&	4.6$_{-2.0}^{+1.5}$	\\
J04520+064	&	 Wolf 1539 	&	 179 	&	 M3.5\,V 	&	3334$_{-54}^{+70}$	&	4.78$_{-0.08}^{+0.13}$	&	0.27$_{-0.18}^{+0.07}$	&	4.8$_{-2.2}^{+1.8}$	\\
J04538--177	&	 GJ 180 	&	 180 	&	 M2.0\,V 	&	3634$_{-40}^{+57}$	&	4.73$_{-0.07}^{+0.05}$	&	0.08$_{-0.12}^{+0.06}$	&	3.4$_{-0.8}^{+1.9}$	\\
J04588+498	&	 BD+49 1280 	&	 181 	&	 M0.0\,V 	&	4008$_{-60}^{+66}$	&	4.62$_{-0.05}^{+0.03}$	&	0.17$_{-0.14}^{+0.10}$	&	5.3$_{-1.4}^{+1.8}$	\\
J05019+011	&	 LP 656-038  	&	 3323 	&	 M4.0\,V 	&	3247$_{-38}^{+37}$	&	4.92$_{-0.05}^{+0.13}$	&	0.30$_{-0.19}^{+0.07}$	&	9.8$_{-1.5}^{+1.6}$	\\
J05019--069	&	 1RXS J050156.7+010845  	&	 ... 	&	 M4.0\,V 	&	3143$_{-59}^{+56}$	&	5.04$_{-0.11}^{+0.10}$	&	0.14$_{-0.25}^{+0.11}$	&	6.3$_{-2.8}^{+1.2}$	\\
J05033--173	&	 LP 776-049 	&	 3325 	&	 M3.0\,V 	&	3385$_{-52}^{+54}$	&	4.84$_{-0.07}^{+0.12}$	&	0.09$_{-0.16}^{+0.07}$	&	5.0$_{-2.4}^{+0.9}$	\\
J05062+046	&	 RX J0506.2+0439  	&	 ... 	&	 M4.0\,V 	&	3204$_{-35}^{+27}$	&	4.86$_{-0.07}^{+0.06}$	&	0.39$_{-0.12}^{+0.09}$	&	28.5$_{-1.2}^{+2.0}$	\\
J05127+196	&	 GJ 192 	&	 192 	&	 M2.0\,V 	&	3610$_{-36}^{+51}$	&	4.72$_{-0.07}^{+0.06}$	&	0.10$_{-0.12}^{+0.07}$	&	3.4$_{-0.9}^{+1.9}$	\\

\noalign{\smallskip}
    \hline
    \end{tabular}
\end{table*}

\subsection{Literature comparison}
\label{lit_comp}

We compared our results to values from the literature, as shown in Fig.~\ref{fig:lit_comparison_CARM}. To increase readability of the plots we present our errorbars in gray. We differentiated between several determination methods to visualize possible biases. 

\subsection*{Effective temperature}

The comparison of our estimated $T_{\rm eff}$ to the following literature works is shown in the top left plot of Fig.~\ref{fig:lit_comparison_CARM}.
Synthetic model fits were performed by Pass19. As mentioned above, \cite{GaidosMann2014} and \cite{Mann2015} derived $T_{\rm eff}$ from visual spectra using BT-Settl model fits. Our values are mostly consistent with these works within their errors. 
Spectral indices and pEWs were used by \cite{RojasAyala2012}, \cite{Maldonado2015}, and \cite{Terrien2015}. The latter study determined three different values of $T_{\rm eff}$ and [M/H] in the $J$, $H$, and $K$ bands. 
We took the values from the $K$ band for comparison, because they report those values to be the most reliable. 
\cite{GaidosMann2014} calculated spectral curvature indices in the near-infrared for stars without visual spectra. In comparison, values from \cite{RojasAyala2012} and \cite{GaidosMann2014} agree well with our results within their errors. We find no correlation with $K$-band $T_{\rm eff}$ from \cite{Terrien2015}, which basically form a straight line around 3300\,K. 

\cite{Houdebine2019} determined $T_{\rm eff}$ using photometric colors. Their results are fairly consistent with our estimates in the temperature range below 3800\,K, but tend to be considerably lower for higher temperatures.
In general, there seems to be no bias regarding different determination methods, since all results follow the same pattern \citep[except for][]{Terrien2015}. 

There is one significant outlier, which is marked with a purple circle. This is \object{GJ~87}, for which we estimated a $T_{\rm eff}$ of 3990$_{-68}^{+106}$\,K, whereas several other works report temperatures 345\,K cooler temperature on average (Pass19: 3605$ \pm 54$\,K; \citealt{Maldonado2015}: 3562$ \pm 68$\,K; \citealt{GaidosMann2014}: 3783$ \pm 94$\,K; \citealt{Mann2015}: 3638$ \pm 62$~K; \cite{Houdebine2019}: 3645$ \pm 33$\,K). However, $\log{g}$, [M/H], and $v \sin{i}$ (see below) are consistent with literature values within their errors. From the spectral type (M1.5\,V) and its nearly solar metallicity (+0.07\,dex), a temperature of around 3700\,K would be more fitting \citep[see Fig.~8 in][]{Passegger2018}, as reported in the literature. At this point we find no explanation for the large deviation in temperature, since this star is not young (e.g., Pass19 assumed 5\,Gyr) and shows no magnetic activity.

Unfortunately, there are only three stars that we have in common with \cite{Antoniadis2020}, who estimated their stellar parameters with the ML tool {\tt ODUSSEAS}. For these stars we derived about 150\,K higher temperatures than with {\tt ODUSSEAS}. Also other literature values for these stars show higher temperatures and match better with our results. 

\subsection*{Surface gravity}
For $\log{g}$ (Fig.~\ref{fig:lit_comparison_CARM}, top right), results from synthetic model fits come from Pass19. 
To obtain $\log{g}$ values for the other literature results, we calculated $\log{g}$ from the masses and radii the respective studies provide. \cite{Mann2015} determined the stellar mass from a mass-luminosity relation \citep{Delfosse2000}, and the stellar radius from employing the Stefan-Boltzmann law with $T_{\rm eff}$ derived from BT-Settl fits and spectroscopically measured bolometric fluxes. \cite{GaidosMann2014} derived stellar masses in the same way as \cite{Mann2015}. For the stellar radius they used the radius-temperature relationship from \cite{Mann2013b}. 
\cite{Maldonado2015} determined stellar masses from empirical photometric relations and stellar radii from a mass-radius relationship involving interferometric measurements and eclipsing binaries. 
Their results are marked as ``interferometric'' in the plot. 
Although most values group along the 1:1 correlation and coincide with literature within their errors, we tentatively estimated lower $\log{g}$ compared to literature. 

\subsection*{Metallicity}
Since our [M/H] results directly translate into identical [Fe/H] values we can compare these with literature [Fe/H] results. Conversely, literature values of [Fe/H] are often interpreted as a proxy for [M/H].
The different measurements are distinguished in the bottom left plot of Fig.~\ref{fig:lit_comparison_CARM}. 
As mentioned in Section~\ref{introduction}, spectral indices, pEWs, and relations of atomic line strength to determine metallicity were used by \cite{RojasAyala2012}, \cite{Maldonado2015}, \cite{Mann2015}, and \cite{GaidosMann2014}. Besides, \cite{Terrien2015} used both pEWs and spectral indices. 
\cite{Dittmann2016} derived [Fe/H] from color-magnitude metallicity relations.
From these studies, only \cite{RojasAyala2012}, \cite{Terrien2015}, and Pass19 provide [M/H], while all others claim [Fe/H].

All our metallicities seem to be constrained to the range 0.0\,dex < [M/H] < +0.4\,dex, in contrast to literature values, which lie in a wider range between --0.4\,dex and +0.5\,dex. 
Ignoring data points from Pass19 (i.e., ``fit [M/H]''), all other literature values are more metal-poor compared to our estimations or, equivalently, our metallicities are too high.
Similar to $T_{\rm eff}$, the three points of \cite{Antoniadis2020} lie in the lower range, presenting metallicities lower than ours and other literature, which might indicate that they tentatively derived too low values.

\subsection*{Rotational velocity}
\begin{figure*}[!ht]
  \centering
  \includegraphics[width=0.49\linewidth]{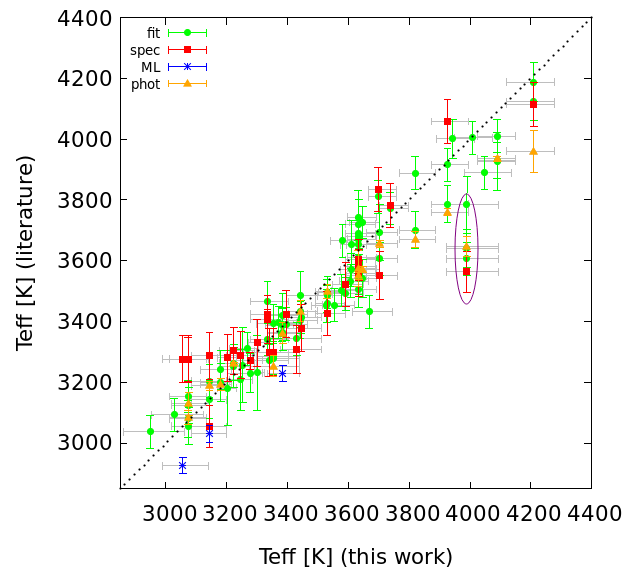}\hfil
  \includegraphics[width=0.49\linewidth]{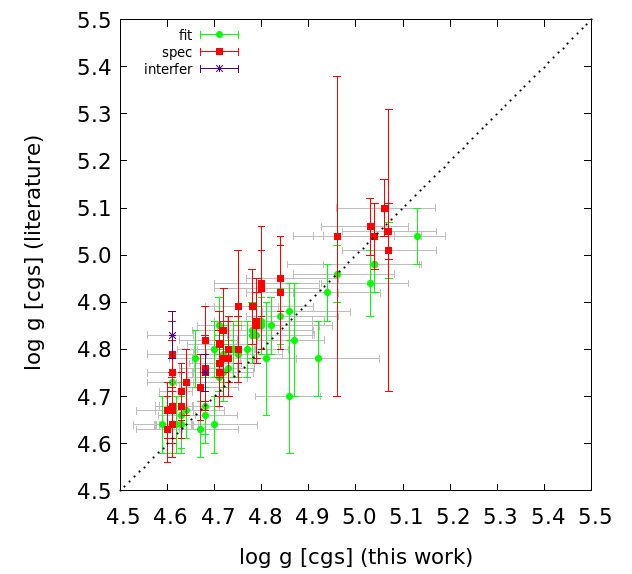}\par\medskip
  \includegraphics[width=0.49\linewidth]{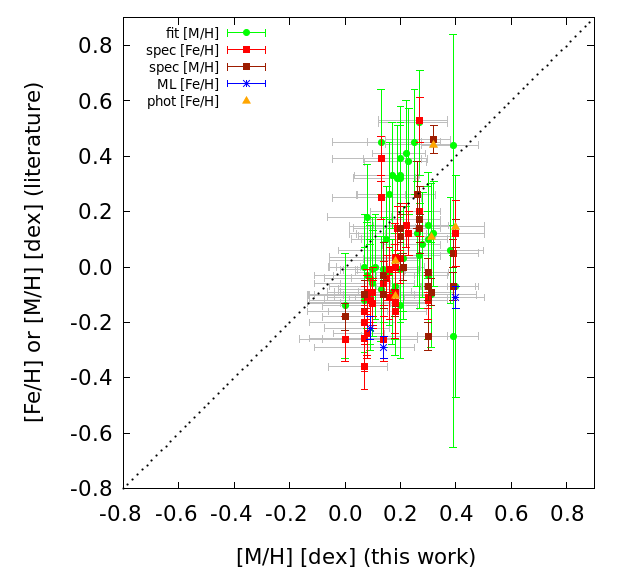}\hfil
  \includegraphics[width=0.49\linewidth]{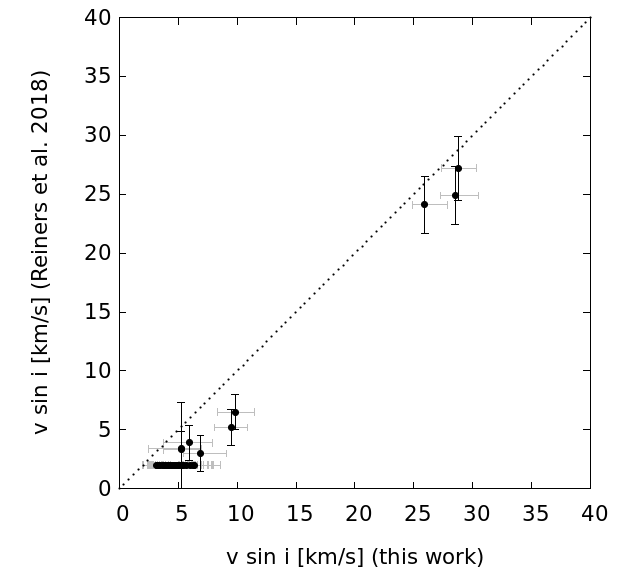}  
\caption{Comparison of the results of this work with values from the literature. For better readability, the errorbars of this work are plotted in gray. Different symbols mark different determination methods of literature values. The ellipse in the top left panel marks the outlier star \object{GJ~87} (J02123+035). Legend: fit (synthetic model fit): \cite{Passegger2019}, \cite{GaidosMann2014} (Teff VIS), \cite{Mann2015} (Teff); spec (pEWs, spectral indices, spectroscopic relations): \cite{RojasAyala2012,Maldonado2015}, \cite{GaidosMann2014} (Teff NIR, log g, [Fe/H]), \cite{Mann2015} (log g, [Fe/H]), \cite{Terrien2015}; ML: \cite{Antoniadis2020}; phot (photometric): \cite{Dittmann2016,Houdebine2019}; interfer (interferometry): \cite{Maldonado2015}}
\label{fig:lit_comparison_CARM}
\end{figure*}

We compared our derived $v \sin{i}$ values with those from \cite{Reiners2018a}, who determined $v \sin{i}$ from CARMENES spectra as well. 
They used a cross-correlation method from several lines in spectral orders of low telluric contamination and high-S/N and provided errors only for $v \sin{i} >$ 2.0\,km\,s$^{-1}$, which is the lower detection limit with the CARMENES VIS spectral resolution.
For our sample, these orders are located at wavelengths $\lambda <$\,6850\,{\AA}. This might explain the differences that we get when we compare $v \sin{i}$. The comparison in the bottom right plot of Fig.~\ref{fig:lit_comparison_CARM} shows that our DL approach slightly overestimates $v \sin{i}$. However, the values are mostly consistent within their errors. 
There are several mechanism that can contribute to the broadening of spectral lines, such as magnetic activity, pressure broadening, and stellar rotation. 
For the first two the strength of the broadening depends on the properties of the considered atomic species.
Only stellar rotation affects all lines in the same way, so it is appropriate to derive $v \sin{i}$ from several lines individually instead of only one small wavelength chunk, as it was done in this work. However, our focus lies toward stellar parameters $T_{\rm eff}$, $\log{g}$, and [M/H].

\subsection{Uncertainties, errors, and the synthetic gap}

Error estimation is almost as challenging as stellar parameter determination itself. There are several ways that authors use to quantify errors in their works. 
In Pass19 the uncertainties were calculated by measuring the standard deviations of the differences between input and output stellar parameters for 1400 noise perturbed synthetic spectra. These uncertainties represent measurement errors of the method itself, but do not take into account the effects of the synthetic gap. For pEW methods, such as \cite{Maldonado2015}, the measurement error in the stellar parameters is directly derived from uncertainties in the pEWs.

\cite{Mann2015} discussed different sources of uncertainties, taking into account measurement errors and errors from calibrations and photometric zero points.
In other works \citep[e.g.,][]{Houdebine2019}, the error is stated as the difference between their results and the literature, which cannot be considered as a measurement or a systematic error. 

\cite{Antoniadis2020} estimated their uncertainties by including the intrinsic uncertainties of the HARPS reference dataset, since these were used in the training process of the ML algorithm and, therefore, introduced noise. This approach can be seen as an attempt at error propagation and measured the influence on the predicted stellar parameters, meaning a unique global standard deviation for all the estimated parameters. 
\cite{Fabbro2018} estimated the prediction error for each stellar parameter by adding in quadrature statistical and intrinsic errors. The statistical error was computed from an error spectrum, giving an approximation of the error on the predicted output stellar parameter. The intrinsic error was empirically measured from a synthetic test sample. 

What a real measurement should attempt to quantify is not only the measurement error but also any type of systematic error. In the case of spectral analyses, one source of systematic errors are the uncertainties inherent in the use of model atmospheres and the derived synthetic spectra. This becomes evident when comparing results of the post-2012 analyses cited in the introduction with those from the decades prior to the BT-Settl models. Important studies based on the previous model generations are, for example, \citet{Leggett2001, Leggett2000} for the AMES-Dusty models \citep{Allard2001}, \citet{Gizis1997} for the NextGen grid \citep{Hauschildt1999}, \citet{Leggett1996} and \citet{Jones1996} for the so called base grid \citep{AllardHauschildt1995}, and \citet{Kirkpatrick1993} for the initial generation of this line of models \citep{Allard1990}. Typically, the cited errors are smaller than the changes that happen when changing the set of model atmospheres. Obviously, the models have improved over the years, both in quality and the amount of details in the implemented physics. However, it is not guaranteed that future improvements in the models will not change the results significantly.

In this work, the synthetic gap gives us a valuable hint on the systematic error. Since it tells us how far away the models are from the observed spectra, the synthetic gap can serve two purposes. The obvious one is a call to improve the model atmospheres and synthetic spectra. Currently, we know which spectral chunks performed the best, that is that they show the smallest MSE during training. In the future we will investigate the variation of the synthetic gap over all the wavelength regions investigated by \citet{Passegger2018} and Pass19, (see Table~\ref{tab:DAT_WIN}), but this is beyond the scope of this paper. Identifying the reasons why some regions can be fitted well, yet showing a large synthetic gap, will point toward potential areas of improvement in the synthetic spectra.
The other aspect is a warning toward those who use models: No synthetic spectrum is perfect even if a $\chi^2$-fit produces excellent agreement.

Another source of uncertainty is using a too small sample of reference stars for training. For example, \citet{Antoniadis2020} used only 65 stars in the training set. 
As discussed by \cite{Fabbro2018}, it is necessary for a deep neural network to have a large training set that spans over the whole stellar parameter range. If there were only a few spectra available for a certain region of the parameter space, such as low temperatures, this would translate into less accurate estimations for stars in this region. This lack of information could become important when using observed spectra as a training set, since there might not be enough observations available to span the whole parameter space. 

Furthermore, a possible misplacement of the continuum, especially for late-type stars, should be taken into account when estimating uncertainties. In our case, we used rather small wavelength ranges of a few 10~\AA{}, and treated the continuum of the synthetic and observed spectra in the same way. As can be seen in Fig.~\ref{fig:window_chunks}, the GISIC algorithm did a good job in normalizing both spectra within this wavelength range. Therefore, we expect the uncertainty coming from a possible continuum misplacement of either spectrum to be negligible. However, because of the way we estimated our errors, those kind of uncertainties were already included. 


\section{Summary and conclusions}
\label{Summary}

We present a deep learning neural network technique to estimate the stellar parameters $T_{\rm eff}$, $\log{g}$, [M/H], and $v \sin{i}$ for M dwarfs from high-S/N, high-resolution, optical, and near-infrared spectroscopy. The DL models were trained with PHOENIX-ACES synthetic spectra, which has the advantage that it is possible to generate a sufficient number of spectra with known stellar parameters. We investigated different architectures and analyzed different spectral windows, which showed only negligible effects on the estimated stellar parameters. We find that all DL models produced only small training and validation errors, meaning that the DL models are able to estimate stellar parameters from synthetic spectra with high precision and accuracy.
In other words, the information content of synthetic spectra (the determination of which is a fundamental astrophysical question) is sufficient to determine the four stellar parameters and is independent of the considered spectral windows.

After constraining the synthetic grid to the M-dwarf parameter space using PARSEC evolutionary models, we trained new DL models on one spectral window and tested their performance on 50 high-resolution CARMENES spectra. Although our results are in good agreement with the literature in most cases (especially for $T_{\rm eff}$ and $\log{g}$), significant deviations are found for the metallicity. We attribute those deviations to the synthetic gap, the difference of spectral feature distribution between synthetic and observed spectra.

To avoid uncertainties introduced by the synthetic gap, we conclude that it seems more practical to use observed spectra with known stellar parameters for training. However, the question arises where those ``known'' parameters come from, since they have to be derived by some method as well, which again introduces uncertainties. Also, it might be difficult to find a sufficient number of observed spectra covering the whole parameter range to properly train an accurate DL model. 

A more detailed study is necessary to quantify the effect of the synthetic gap on stellar parameter determination and get to the bottom of its sources. For this study the main aim was to validate the DL method, which delivers satisfying results using synthetic models. However, shortcomings are identified when applying the trained DL models to observed spectra. Therefore, we will investigate the synthetic gap and its effect on [M/H] in more detail in a following study. We will also add noise to the synthetic spectra, which will help to simulate a more realistic setup in the training step, however, this will have no significant influence on the synthetic gap.
In summary, we present a method to derive stellar parameters with DL models trained on synthetic spectra and estimated the uncertainty related to the synthetic gap. This work should also be seen as a word of caution toward scientists employing synthetic spectra uncritically in their work, since even apparently perfect fits do not necessarily provide perfect results, which should be accounted for with larger error estimates. 

%




\begin{acknowledgements}
We thank an anonymous referee for helpful comments that improved the quality of this paper.
CARMENES is an instrument for the Centro Astron\'omico Hispano-Alem\'an de Calar Alto (CAHA, Almer\'ia, Spain). 
CARMENES is funded by the German Max-Planck-Gesellschaft (MPG), the Spanish Consejo Superior de Investigaciones Cient\'ificas (CSIC), European Regional Development Fund (ERDF) through projects FICTS-2011-02, ICTS-2017-07-CAHA-4, and CAHA16-CE-3978, 
and the members of the CARMENES Consortium (Max-Planck-Institut f\"ur Astronomie, Instituto de Astrof\'isica de Andaluc\'ia, Landessternwarte K\"onigstuhl, Institut de Ci\`encies de l'Espai, Insitut f\"ur Astrophysik G\"ottingen, Universidad Complutense de Madrid, Th\"uringer Landessternwarte Tautenburg, Instituto de Astrof\'isica de Canarias, Hamburger Sternwarte, Centro de Astrobiolog\'ia and Centro Astron\'omico Hispano-Alem\'an), with additional contributions by the Spanish Ministry of Economy, the German Science Foundation through the Major Research Instrumentation Programme and DFG Research Unit FOR2544 ``Blue Planets around Red Stars'', the Klaus Tschira Stiftung, the states of Baden-W\"urttemberg and Niedersachsen, and by the Junta de Andaluc\'ia.
We acknowledge financial support from NASA through grant NNX17AG24G, 
the Agencia Estatal de Investigaci\'on of the Ministerio de Ciencia through fellowship FPU15/01476, Innovaci\'on y Universidades and the ERDF through projects
PID2019-109522GB-C51/2/3/4, 
AYA2016-79425-C3-1/2/3-P	
and AYA2018-84089,  		
the Funda\c{c}\~{a}o para a Ci\^{e}ncia e a Tecnologia through and ERDF through grants UID/FIS/04434/2019, UIDB/04434/2020 and UIDP/04434/2020, PTDC/FIS-AST/28953/2017, and COMPETE2020 - Programa Operacional Competitividade e Internacionaliza\c{c}\~{a}o POCI-01-0145-FEDER-028953.

\end{acknowledgements}
\bibliographystyle{aa}  
\bibliography{CARM_DL.bib}

\begin{thebibliography}{131}
\expandafter\ifx\csname natexlab\endcsname\relax\def\natexlab#1{#1}\fi

\bibitem[{Abadi(2015)}]{Abadi2015}
Abadi, M., e.~a. 2015, TensorFlow: Large-Scale Machine Learning on
  Heterogeneous Systems, \url{https://github.com/tensorflow/tensorflow},
  accessed: 2020-02-07

\bibitem[{Abraham {et~al.}(2018)Abraham, Aniyan, Kembhavi, Philip, \&
  Vaghmare}]{Abraham2018}
Abraham, S., Aniyan, A.~K., Kembhavi, A.~K., Philip, N.~S., \& Vaghmare, K.
  2018, Monthly Notices of the Royal Astronomical Society, 477, 894

\bibitem[{{Allard}(1990)}]{Allard1990}
{Allard}, F. 1990, PhD thesis, Centre de Recherche Astrophysique de Lyon

\bibitem[{{Allard} \& {Hauschildt}(1995)}]{AllardHauschildt1995}
{Allard}, F. \& {Hauschildt}, P.~H. 1995, \apj, 445, 433

\bibitem[{{Allard} {et~al.}(2001){Allard}, {Hauschildt}, {Alexander},
  {Tamanai}, \& {Schweitzer}}]{Allard2001}
{Allard}, F., {Hauschildt}, P.~H., {Alexander}, D.~R., {Tamanai}, A., \&
  {Schweitzer}, A. 2001, \apj, 556, 357

\bibitem[{{Allard} {et~al.}(2011){Allard}, {Homeier}, \&
  {Freytag}}]{Allard2011}
{Allard}, F., {Homeier}, D., \& {Freytag}, B. 2011, in Astronomical Society of
  the Pacific Conference Series, Vol. 448, 16th Cambridge Workshop on Cool
  Stars, Stellar Systems, and the Sun, ed. C.~{Johns-Krull}, M.~K. {Browning},
  \& A.~A. {West}, 91

\bibitem[{{Allard} {et~al.}(2012){Allard}, {Homeier}, \&
  {Freytag}}]{Allard2012}
{Allard}, F., {Homeier}, D., \& {Freytag}, B. 2012, Philosophical Transactions
  of the Royal Society of London Series A, 370, 2765

\bibitem[{{Allard} {et~al.}(2013){Allard}, {Homeier}, {Freytag},
  {Schaffenberger}, {}, \& {Rajpurohit}}]{Allard2013}
{Allard}, F., {Homeier}, D., {Freytag}, B., {et~al.} 2013, Memorie della
  Societa Astronomica Italiana Supplementi, 24, 128

\bibitem[{{Alvarez} \& {Plez}(1998)}]{AlvarezPlez1998}
{Alvarez}, R. \& {Plez}, B. 1998, \aap, 330, 1109

\bibitem[{Anthony \& Bartlett(2009)}]{anthony2009neural}
Anthony, M. \& Bartlett, P.~L. 2009, Neural network learning: Theoretical
  foundations (cambridge university press)

\bibitem[{{Antoniadis-Karnavas} {et~al.}(2020){Antoniadis-Karnavas}, {Sousa},
  {Delgado-Mena}, {Santos}, {Teixeira}, \& {Neves}}]{Antoniadis2020}
{Antoniadis-Karnavas}, A., {Sousa}, S.~G., {Delgado-Mena}, E., {et~al.} 2020,
  \aap, 636, A9

\bibitem[{{Bailer-Jones} {et~al.}(1997){Bailer-Jones}, {Irwin}, {Gilmore}, \&
  {von Hippel}}]{Bailer-Jones1997}
{Bailer-Jones}, C. A.~L., {Irwin}, M., {Gilmore}, G., \& {von Hippel}, T. 1997,
  \mnras, 292, 157

\bibitem[{{Bean} {et~al.}(2006){Bean}, {Benedict}, \& {Endl}}]{Bean2006}
{Bean}, J.~L., {Benedict}, G.~F., \& {Endl}, M. 2006, \apjl, 653, L65

\bibitem[{Birky {et~al.}(2020)Birky, Hogg, Mann, \& Burgasser}]{Birky2020}
Birky, J., Hogg, D.~W., Mann, A.~W., \& Burgasser, A. 2020, The Astrophysical
  Journal, 892, 31

\bibitem[{{Birky} {et~al.}(2017){Birky}, {Aganze}, {Burgasser}, {Theissen},
  {Schmidt}, {Teske}, {Stassun}, {Bird}, \& {UCSD FAST Team}}]{Birky2017}
{Birky}, J.~L., {Aganze}, C., {Burgasser}, A.~J., {et~al.} 2017, in American
  Astronomical Society Meeting Abstracts, Vol. 229, American Astronomical
  Society Meeting Abstracts \#229, 240.18

\bibitem[{{Bonfils} {et~al.}(2005){Bonfils}, {Delfosse}, {Udry}, {Santos},
  {Forveille}, \& {S{\'e}gransan}}]{Bonfils2005}
{Bonfils}, X., {Delfosse}, X., {Udry}, S., {et~al.} 2005, \aap, 442, 635

\bibitem[{{Boyajian} {et~al.}(2012){Boyajian}, {von Braun}, {van Belle},
  {McAlister}, {ten Brummelaar}, {Kane}, {Muirhead}, {Jones}, {White},
  {Schaefer}, {Ciardi}, {Henry}, {L{\'o}pez-Morales}, {Ridgway}, {Gies}, {Jao},
  {Rojas-Ayala}, {Parks}, {Sturmann}, {Sturmann}, {Turner}, {Farrington},
  {Goldfinger}, \& {Berger}}]{Boyajian2012}
{Boyajian}, T.~S., {von Braun}, K., {van Belle}, G., {et~al.} 2012, \apj, 757,
  112

\bibitem[{{Bressan} {et~al.}(2012){Bressan}, {Marigo}, {Girardi}, {Salasnich},
  {Dal Cero}, {Rubele}, \& {Nanni}}]{Bressan2012}
{Bressan}, A., {Marigo}, P., {Girardi}, L., {et~al.} 2012, \mnras, 427, 127

\bibitem[{{Caballero} {et~al.}(2016){Caballero}, {Gu{\`a}rdia}, {L{\'o}pez del
  Fresno}, {Zechmeister}, {de Juan}, {Alonso-Floriano}, {Amado}, {Colom{\'e}},
  {Cort{\'e}s-Contreras}, {Garc{\'{\i}}a-Piquer}, {Gesa}, {de Guindos},
  {Hagen}, {Helmling}, {Hern{\'a}ndez Casta{\~n}o}, {K{\"u}rster},
  {L{\'o}pez-Santiago}, {Montes}, {Morales Mu{\~n}oz}, {Pavlov}, {Quirrenbach},
  {Reiners}, {Ribas}, {Seifert}, \& {Solano}}]{Caballero2016}
{Caballero}, J.~A., {Gu{\`a}rdia}, J., {L{\'o}pez del Fresno}, M., {et~al.}
  2016, in \procspie, Vol. 9910, Observatory Operations: Strategies, Processes,
  and Systems VI, 99100E

\bibitem[{{Casagrande} {et~al.}(2008){Casagrande}, {Flynn}, \&
  {Bessell}}]{Casagrande2008}
{Casagrande}, L., {Flynn}, C., \& {Bessell}, M. 2008, \mnras, 389, 585

\bibitem[{{Casey} {et~al.}(2016){Casey}, {Hogg}, {Ness}, {Rix}, {Ho}, \&
  {Gilmore}}]{Casey2016}
{Casey}, A.~R., {Hogg}, D.~W., {Ness}, M., {et~al.} 2016, arXiv e-prints,
  arXiv:1603.03040

\bibitem[{{Chen} {et~al.}(2015){Chen}, {Bressan}, {Girardi}, {Marigo}, {Kong},
  \& {Lanza}}]{Chen2015}
{Chen}, Y., {Bressan}, A., {Girardi}, L., {et~al.} 2015, \mnras, 452, 1068

\bibitem[{{Chen} {et~al.}(2014){Chen}, {Girardi}, {Bressan}, {Marigo},
  {Barbieri}, \& {Kong}}]{Chen2014}
{Chen}, Y., {Girardi}, L., {Bressan}, A., {et~al.} 2014, \mnras, 444, 2525

\bibitem[{{Chollet}(2015)}]{Chollet2015}
{Chollet}, F. 2015, KERAS, \url{https://github.com/keras-team/keras}, accessed:
  2020-02-07

\bibitem[{{Czesla} {et~al.}(2019){Czesla}, {Schr{\"o}ter}, {Schneider},
  {Huber}, {Pfeifer}, {Andreasen}, \& {Zechmeister}}]{pya}
{Czesla}, S., {Schr{\"o}ter}, S., {Schneider}, C.~P., {et~al.} 2019, {PyA:
  Python astronomy-related packages}

\bibitem[{{Delfosse} {et~al.}(2000){Delfosse}, {Forveille}, {S{\'e}gransan},
  {Beuzit}, {Udry}, {Perrier}, \& {Mayor}}]{Delfosse2000}
{Delfosse}, X., {Forveille}, T., {S{\'e}gransan}, D., {et~al.} 2000, \aap, 364,
  217

\bibitem[{{Demory} {et~al.}(2009){Demory}, {S{\'e}gransan}, {Forveille},
  {Queloz}, {Beuzit}, {Delfosse}, {di Folco}, {Kervella}, {Le Bouquin},
  {Perrier}, {Benisty}, {Duvert}, {Hofmann}, {Lopez}, \& {Petrov}}]{Demory2009}
{Demory}, B.~O., {S{\'e}gransan}, D., {Forveille}, T., {et~al.} 2009, \aap,
  505, 205

\bibitem[{{Dhital} {et~al.}(2012){Dhital}, {West}, {Stassun}, {Bochanski},
  {Massey}, \& {Bastien}}]{Dhital2012}
{Dhital}, S., {West}, A.~A., {Stassun}, K.~G., {et~al.} 2012, \aj, 143, 67

\bibitem[{Dieleman {et~al.}(2015)Dieleman, Willett, \& Dambre}]{Dieleman2015}
Dieleman, S., Willett, K.~W., \& Dambre, J. 2015, Monthly Notices of the Royal
  Astronomical Society, 450, 1441

\bibitem[{Dittmann {et~al.}(2016)Dittmann, Irwin, Charbonneau, \&
  Newton}]{Dittmann2016}
Dittmann, J.~A., Irwin, J.~M., Charbonneau, D., \& Newton, E.~R. 2016, The
  Astrophysical Journal, 818, 153

\bibitem[{{Fabbro} {et~al.}(2018){Fabbro}, {Venn}, {O'Briain}, {Bialek},
  {Kielty}, {Jahandar}, \& {Monty}}]{Fabbro2018}
{Fabbro}, S., {Venn}, K.~A., {O'Briain}, T., {et~al.} 2018, \mnras, 475, 2978

\bibitem[{Figueira {et~al.}(2016)Figueira, Adibekyan, Oshagh, Neal,
  Rojas-Ayala, Lovis, Melo, Pepe, Santos, \& Tsantaki}]{Figueira2016}
Figueira, P., Adibekyan, V.~Z., Oshagh, M., {et~al.} 2016, Astronomy {\&}
  Astrophysics, 586, A101

\bibitem[{Freedman \& Diaconis(1981)}]{FreedmanDiaconis1981}
Freedman, D. \& Diaconis, P. 1981, Zeitschrift f\"ur Wahrscheinlichkeitstheorie
  und Verwandte Gebiete, 57, 453

\bibitem[{{Gaidos} \& {Mann}(2014)}]{GaidosMann2014}
{Gaidos}, E. \& {Mann}, A.~W. 2014, \apj, 791, 54

\bibitem[{{Gizis}(1997)}]{Gizis1997}
{Gizis}, J.~E. 1997, \aj, 113, 806

\bibitem[{Gong \& Ordieres-Mer{\'{e}}(2016)}]{Gong2016}
Gong, B. \& Ordieres-Mer{\'{e}}, J. 2016, Environmental Modelling {\&}
  Software, 84, 290

\bibitem[{Gonz{\'{a}}lez-Marcos {et~al.}(2013)Gonz{\'{a}}lez-Marcos,
  Ordieres-Mer{\'{e}}, Alba-El{\'{\i}}as, de~Pis{\'{o}}n, \&
  Castej{\'{o}}n-Limas}]{GonzalezMarcos2013}
Gonz{\'{a}}lez-Marcos, A., Ordieres-Mer{\'{e}}, J., Alba-El{\'{\i}}as, F.,
  de~Pis{\'{o}}n, F. J.~M., \& Castej{\'{o}}n-Limas, M. 2013, Ironmaking {\&}
  Steelmaking, 41, 262

\bibitem[{{Gonz{\'a}lez-Marcos} {et~al.}(2017){Gonz{\'a}lez-Marcos}, {Sarro},
  {Ordieres-Mer{\'e}}, \& {Bello-Garc{\'\i}a}}]{GonzalezMarcos2017}
{Gonz{\'a}lez-Marcos}, A., {Sarro}, L.~M., {Ordieres-Mer{\'e}}, J., \&
  {Bello-Garc{\'\i}a}, A. 2017, \mnras, 465, 4556

\bibitem[{{Gulati} {et~al.}(1994){Gulati}, {Gupta}, {Gothoskar}, \&
  {Khobragade}}]{Gulati1994}
{Gulati}, R.~K., {Gupta}, R., {Gothoskar}, P., \& {Khobragade}, S. 1994, \apj,
  426, 340

\bibitem[{{Gustafsson} {et~al.}(2008){Gustafsson}, {Edvardsson}, {Eriksson},
  {J{\o}rgensen}, {Nordlund}, \& {Plez}}]{Gustafsson2008}
{Gustafsson}, B., {Edvardsson}, B., {Eriksson}, K., {et~al.} 2008, \aap, 486,
  951

\bibitem[{{Hafner} \& {Wehrse}(1994)}]{Hafner1994}
{Hafner}, M. \& {Wehrse}, R. 1994, \aap, 282, 874

\bibitem[{{Hartman} {et~al.}(2015){Hartman}, {Bayliss}, {Brahm}, {Bakos},
  {Mancini}, {Jord{\'a}n}, {Penev}, {Rabus}, {Zhou}, {Butler}, {Espinoza}, {de
  Val-Borro}, {Bhatti}, {Csubry}, {Ciceri}, {Henning}, {Schmidt}, {Arriagada},
  {Shectman}, {Crane}, {Thompson}, {Suc}, {Cs{\'a}k}, {Tan}, {Noyes},
  {L{\'a}z{\'a}r}, {Papp}, \& {S{\'a}ri}}]{Hartman2015}
{Hartman}, J.~D., {Bayliss}, D., {Brahm}, R., {et~al.} 2015, \aj, 149, 166

\bibitem[{{Hauschildt}(1992)}]{Hauschildt1992}
{Hauschildt}, P.~H. 1992, \jqsrt, 47, 433

\bibitem[{{Hauschildt}(1993)}]{Hauschildt1993}
{Hauschildt}, P.~H. 1993, \jqsrt, 50, 301

\bibitem[{{Hauschildt} {et~al.}(1999){Hauschildt}, {Allard}, \&
  {Baron}}]{Hauschildt1999}
{Hauschildt}, P.~H., {Allard}, F., \& {Baron}, E. 1999, \apj, 512, 377

\bibitem[{{Hauschildt} \& {Baron}(1999)}]{HauschildtBaron1999}
{Hauschildt}, P.~H. \& {Baron}, E. 1999, Journal of Computational and Applied
  Mathematics, 109, 41

\bibitem[{{He} \& {Zhao}(2019)}]{He2019}
{He}, W. \& {Zhao}, G. 2019, Research in Astronomy and Astrophysics, 19, 140

\bibitem[{{Hon} {et~al.}(2017){Hon}, {Stello}, \& {Yu}}]{Hon2017}
{Hon}, M., {Stello}, D., \& {Yu}, J. 2017, \mnras, 469, 4578

\bibitem[{Houdebine {et~al.}(2019)Houdebine, Mullan, Doyle, de~La~Vieuville,
  Butler, \& Paletou}]{Houdebine2019}
Houdebine, {\'{E}}.~R., Mullan, D.~J., Doyle, J.~G., {et~al.} 2019, The
  Astronomical Journal, 158, 56

\bibitem[{{Husser} {et~al.}(2013){Husser}, {Wende-von Berg}, {Dreizler},
  {Homeier}, {Reiners}, {Barman}, \& {Hauschildt}}]{Husser2013}
{Husser}, T.-O., {Wende-von Berg}, S., {Dreizler}, S., {et~al.} 2013, \aap,
  553, A6

\bibitem[{{Johnson} \& {Apps}(2009)}]{JohnsonApps2009}
{Johnson}, J.~A. \& {Apps}, K. 2009, \apj, 699, 933

\bibitem[{{Johnson} {et~al.}(2012){Johnson}, {Gazak}, {Apps}, {Muirhead},
  {Crepp}, {Crossfield}, {Boyajian}, {von Braun}, {Rojas-Ayala}, {Howard},
  {Covey}, {Schlawin}, {Hamren}, {Morton}, {Marcy}, \& {Lloyd}}]{Johnson2012}
{Johnson}, J.~A., {Gazak}, J.~Z., {Apps}, K., {et~al.} 2012, \aj, 143, 111

\bibitem[{Johnson~S.G.(2019)}]{libcerf}
Johnson~S.G., Cervellino~A., W.~J. 2019, libcerf, numeric library for complex
  error functions, version 1.13, \url{https://jugit.fz-juelich.de/mlz/libcerf}

\bibitem[{{Jones} {et~al.}(1996){Jones}, {Longmore}, {Allard}, \&
  {Hauschildt}}]{Jones1996}
{Jones}, H. R.~A., {Longmore}, A.~J., {Allard}, F., \& {Hauschildt}, P.~H.
  1996, \mnras, 280, 77

\bibitem[{{Kausch} {et~al.}(2014){Kausch}, {Noll}, {Smette}, {Kimeswenger},
  {Horst}, {Sana}, {Jones}, {Barden}, {Szyszka}, \& {Vinther}}]{Kausch2014}
{Kausch}, W., {Noll}, S., {Smette}, A., {et~al.} 2014, in Astronomical Society
  of the Pacific Conference Series, Vol. 485, Astronomical Data Analysis
  Software and Systems XXIII, ed. N.~{Manset} \& P.~{Forshay}, 403

\bibitem[{{Khata} {et~al.}(2020){Khata}, {Mondal}, {Das}, {Ghosh}, \&
  {Ghosh}}]{Khata2020}
{Khata}, D., {Mondal}, S., {Das}, R., {Ghosh}, S., \& {Ghosh}, S. 2020, \mnras,
  493, 4533

\bibitem[{Kielty {et~al.}(2018)Kielty, Bialek, Fabbro, Venn, O'Briain,
  Jahandar, \& Monty}]{kielty2018starnet}
Kielty, C.~L., Bialek, S., Fabbro, S., {et~al.} 2018, in Software and
  cyberinfrastructure for astronomy v, Vol. 10707, International Society for
  Optics and Photonics, 107072W

\bibitem[{{Kirkpatrick} {et~al.}(1993){Kirkpatrick}, {Kelly}, {Rieke},
  {Liebert}, {Allard}, \& {Wehrse}}]{Kirkpatrick1993}
{Kirkpatrick}, J.~D., {Kelly}, D.~M., {Rieke}, G.~H., {et~al.} 1993, \apj, 402,
  643

\bibitem[{{Kuznetsov} {et~al.}(2019){Kuznetsov}, {del Burgo}, {Pavlenko}, \&
  {Frith}}]{Kuznetsov2019}
{Kuznetsov}, M.~K., {del Burgo}, C., {Pavlenko}, Y.~V., \& {Frith}, J. 2019,
  \apj, 878, 134

\bibitem[{LeCun {et~al.}(2015)LeCun, Bengio, \& Hinton}]{lecun2015deep}
LeCun, Y., Bengio, Y., \& Hinton, G. 2015, nature, 521, 436

\bibitem[{{Leggett} {et~al.}(1996){Leggett}, {Allard}, {Berriman}, {Dahn}, \&
  {Hauschildt}}]{Leggett1996}
{Leggett}, S.~K., {Allard}, F., {Berriman}, G., {Dahn}, C.~C., \& {Hauschildt},
  P.~H. 1996, \apjs, 104, 117

\bibitem[{{Leggett} {et~al.}(2000){Leggett}, {Allard}, {Dahn}, {Hauschildt},
  {Kerr}, \& {Rayner}}]{Leggett2000}
{Leggett}, S.~K., {Allard}, F., {Dahn}, C., {et~al.} 2000, \apj, 535, 965

\bibitem[{{Leggett} {et~al.}(2001){Leggett}, {Allard}, {Geballe}, {Hauschildt},
  \& {Schweitzer}}]{Leggett2001}
{Leggett}, S.~K., {Allard}, F., {Geballe}, T.~R., {Hauschildt}, P.~H., \&
  {Schweitzer}, A. 2001, \apj, 548, 908

\bibitem[{{Leung} \& {Bovy}(2019)}]{Leung2019}
{Leung}, H.~W. \& {Bovy}, J. 2019, \mnras, 483, 3255

\bibitem[{Li {et~al.}(2017)Li, Pan, \& Duan}]{li2017parameterizing}
Li, X.-R., Pan, R.-Y., \& Duan, F.-Q. 2017, Research in Astronomy and
  Astrophysics, 17, 036

\bibitem[{{Mahabal} {et~al.}(2017){Mahabal}, {Sheth}, {Gieseke}, {Pai},
  {Djorgovski}, {Drake}, {Graham}, \& {the CSS/CRTS/PTF
  Collaboration}}]{Mahabal2017}
{Mahabal}, A., {Sheth}, K., {Gieseke}, F., {et~al.} 2017, arXiv e-prints,
  arXiv:1709.06257

\bibitem[{{Maldonado} {et~al.}(2015){Maldonado}, {Affer}, {Micela},
  {Scandariato}, {Damasso}, {Stelzer}, {Barbieri}, {Bedin}, {Biazzo},
  {Bignamini}, {Borsa}, {Claudi}, {Covino}, {Desidera}, {Esposito}, {Gratton},
  {Gonz{\'a}lez Hern{\'a}ndez}, {Lanza}, {Maggio}, {Molinari}, {Pagano},
  {Perger}, {Pillitteri}, {Piotto}, {Poretti}, {Prisinzano}, {Rebolo}, {Ribas},
  {Shkolnik}, {Southworth}, {Sozzetti}, \& {Su{\'a}rez
  Mascare{\~n}o}}]{Maldonado2015}
{Maldonado}, J., {Affer}, L., {Micela}, G., {et~al.} 2015, \aap, 577, A132

\bibitem[{{Mann} {et~al.}(2013{\natexlab{a}}){Mann}, {Brewer}, {Gaidos},
  {L{\'e}pine}, \& {Hilton}}]{Mann2013a}
{Mann}, A.~W., {Brewer}, J.~M., {Gaidos}, E., {L{\'e}pine}, S., \& {Hilton},
  E.~J. 2013{\natexlab{a}}, \aj, 145, 52

\bibitem[{{Mann} {et~al.}(2014){Mann}, {Deacon}, {Gaidos}, {Ansdell}, {Brewer},
  {Liu}, {Magnier}, \& {Aller}}]{Mann2014}
{Mann}, A.~W., {Deacon}, N.~R., {Gaidos}, E., {et~al.} 2014, \aj, 147, 160

\bibitem[{{Mann} {et~al.}(2015){Mann}, {Feiden}, {Gaidos}, {Boyajian}, \& {von
  Braun}}]{Mann2015}
{Mann}, A.~W., {Feiden}, G.~A., {Gaidos}, E., {Boyajian}, T., \& {von Braun},
  K. 2015, \apj, 804, 64

\bibitem[{{Mann} {et~al.}(2013{\natexlab{b}}){Mann}, {Gaidos}, \&
  {Ansdell}}]{Mann2013b}
{Mann}, A.~W., {Gaidos}, E., \& {Ansdell}, M. 2013{\natexlab{b}}, \apj, 779,
  188

\bibitem[{McCarthy \& Hayes(1981)}]{mccarthy1981some}
McCarthy, J. \& Hayes, P.~J. 1981, in Readings in artificial intelligence
  (Elsevier), 431--450

\bibitem[{McInnes {et~al.}(2018)McInnes, Healy, Saul, \&
  Grossberger}]{mcinnes2018umap-software}
McInnes, L., Healy, J., Saul, N., \& Grossberger, L. 2018, The Journal of Open
  Source Software, 3, 861

\bibitem[{Meyer(2017)}]{Meyer2017}
Meyer, M. 2017, PhD thesis, Universit\"{a}t Hamburg, Germany

\bibitem[{Mittal \& Vaishay(2019)}]{Mittal2019}
Mittal, S. \& Vaishay, S. 2019, Journal of Systems Architecture, 99, 101635

\bibitem[{{Montes} {et~al.}(2018){Montes}, {Gonz{\'a}lez-Peinado}, {Tabernero},
  {Caballero}, {Marfil}, {Alonso-Floriano}, {Cort{\'e}s-Contreras},
  {Gonz{\'a}lez Hern{\'a}ndez}, {Klutsch}, \& {Moreno-J{\'o}dar}}]{Montes2018}
{Montes}, D., {Gonz{\'a}lez-Peinado}, R., {Tabernero}, H.~M., {et~al.} 2018,
  \mnras, 479, 1332

\bibitem[{Nagel {et~al.}(2020, submitted)Nagel, Czesla, Kaminski, Zechmeister,
  Tal-Or, Schmitt, Reiners, Quirrenbach, Amado, Caballero, Alacid, Bauer,
  B{\'{e}}jar, Cort{\'{e}}s-Contreras, Dreizler, Hatzes, Jeffers, Kürster,
  Lafarga, Montes, Morales, \& Pedraz}]{Nagel2020}
Nagel, E., Czesla, S., Kaminski, A., {et~al.} 2020, submitted, \aap

\bibitem[{Nemravov{\'{a}} {et~al.}(2016)Nemravov{\'{a}}, Harmanec, Bro{\v{z}},
  Vokrouhlick{\'{y}}, Mourard, Hummel, Cameron, Matthews, Bolton,
  Bo{\v{z}}i{\'{c}}, Chini, Dembsky, Engle, Farrington, Grunhut, Guenther,
  Guinan, Kor{\v{c}}{\'{a}}kov{\'{a}}, Koubsk{\'{y}},
  K{\v{r}}{\'{\i}}{\v{c}}ek, Kuschnig, Mayer, McCook, Moffat, Nardetto,
  Pr{\v{s}}a, Ribeiro, Rowe, Rucinski, {\v{S}}koda, {\v{S}}lechta, Tallon-Bosc,
  Votruba, Weiss, Wolf, Zasche, \& Zavala}]{Nemravov2016}
Nemravov{\'{a}}, J.~A., Harmanec, P., Bro{\v{z}}, M., {et~al.} 2016, Astronomy
  {\&} Astrophysics, 594, A55

\bibitem[{{Ness} {et~al.}(2015){Ness}, {Hogg}, {Rix}, {Ho}, \&
  {Zasowski}}]{Ness2015}
{Ness}, M., {Hogg}, D.~W., {Rix}, H.~W., {Ho}, A. Y.~Q., \& {Zasowski}, G.
  2015, \apj, 808, 16

\bibitem[{{Neves} {et~al.}(2012){Neves}, {Bonfils}, {Santos}, {Delfosse},
  {Forveille}, {Allard}, {Nat{\'a}rio}, {Fernand es}, \& {Udry}}]{Neves2012}
{Neves}, V., {Bonfils}, X., {Santos}, N.~C., {et~al.} 2012, \aap, 538, A25

\bibitem[{{Neves} {et~al.}(2013){Neves}, {Bonfils}, {Santos}, {Delfosse},
  {Forveille}, {Allard}, \& {Udry}}]{Neves2013}
{Neves}, V., {Bonfils}, X., {Santos}, N.~C., {et~al.} 2013, \aap, 551, A36

\bibitem[{{Neves} {et~al.}(2014){Neves}, {Bonfils}, {Santos}, {Delfosse},
  {Forveille}, {Allard}, \& {Udry}}]{Neves2014}
{Neves}, V., {Bonfils}, X., {Santos}, N.~C., {et~al.} 2014, \aap, 568, A121

\bibitem[{{Newton} {et~al.}(2014){Newton}, {Charbonneau}, {Irwin},
  {Berta-Thompson}, {Rojas-Ayala}, {Covey}, \& {Lloyd}}]{Newton2014}
{Newton}, E.~R., {Charbonneau}, D., {Irwin}, J., {et~al.} 2014, \aj, 147, 20

\bibitem[{{Newton} {et~al.}(2015){Newton}, {Charbonneau}, {Irwin}, \&
  {Mann}}]{Newton2015}
{Newton}, E.~R., {Charbonneau}, D., {Irwin}, J., \& {Mann}, A.~W. 2015, \apj,
  800, 85

\bibitem[{{{\"O}nehag} {et~al.}(2012){{\"O}nehag}, {Heiter}, {Gustafsson},
  {Piskunov}, {Plez}, \& {Reiners}}]{Onehag2012}
{{\"O}nehag}, A., {Heiter}, U., {Gustafsson}, B., {et~al.} 2012, \aap, 542, A33

\bibitem[{{Paletou} {et~al.}(2015){Paletou}, {Gebran}, {Houdebine}, \&
  {Watson}}]{Paletou2015}
{Paletou}, F., {Gebran}, M., {Houdebine}, E.~R., \& {Watson}, V. 2015, \aap,
  580, A78

\bibitem[{Parzen(1962)}]{Parzen1962}
Parzen, E. 1962, The Annals of Mathematical Statistics, 33, 1065

\bibitem[{{Passegger} {et~al.}(2018){Passegger}, {Reiners}, {Jeffers},
  {Wende-von Berg}, {Sch{\"o}fer}, {Caballero}, {Schweitzer}, {Amado},
  {B{\'e}jar}, {Cort{\'e}s-Contreras}, {Hatzes}, {K{\"u}rster}, {Montes},
  {Pedraz}, {Quirrenbach}, {Ribas}, \& {Seifert}}]{Passegger2018}
{Passegger}, V.~M., {Reiners}, A., {Jeffers}, S.~V., {et~al.} 2018, \aap, 615,
  A6

\bibitem[{{Passegger} {et~al.}(2019){Passegger}, {Schweitzer}, {Shulyak},
  {Nagel}, {Hauschildt}, {Reiners}, {Amado}, {Caballero},
  {Cort{\'e}s-Contreras}, {Dom{\'\i}nguez-Fern{\'a}ndez}, {Quirrenbach},
  {Ribas}, {Azzaro}, {Anglada-Escud{\'e}}, {Bauer}, {B{\'e}jar}, {Dreizler},
  {Guenther}, {Henning}, {Jeffers}, {Kaminski}, {K{\"u}rster}, {Lafarga},
  {Mart{\'\i}n}, {Montes}, {Morales}, {Schmitt}, \&
  {Zechmeister}}]{Passegger2019}
{Passegger}, V.~M., {Schweitzer}, A., {Shulyak}, D., {et~al.} 2019, \aap, 627,
  A161

\bibitem[{Petersen \& Voigtlaender(2018)}]{petersen2018optimal}
Petersen, P. \& Voigtlaender, F. 2018, Neural Networks, 108, 296

\bibitem[{{Plez}(2012)}]{Plez2012}
{Plez}, B. 2012, {Turbospectrum: Code for spectral synthesis}

\bibitem[{{Quirrenbach} {et~al.}(2018){Quirrenbach}, {Amado}, {Ribas},
  {Reiners}, {Caballero}, {Seifert}, {Aceituno}, {Azzaro}, {Baroch}, {Barrado},
  \& et~al.}]{Quirrenbach2018}
{Quirrenbach}, A., {Amado}, P.~J., {Ribas}, I., {et~al.} 2018, in Society of
  Photo-Optical Instrumentation Engineers (SPIE) Conference Series, Vol. 10702,
  Society of Photo-Optical Instrumentation Engineers (SPIE) Conference Series,
  107020W

\bibitem[{{Rajpurohit} {et~al.}(2018){Rajpurohit}, {Allard}, {Rajpurohit},
  {Sharma}, {Teixeira}, {Mousis}, \& {Kamlesh}}]{Rajpurohit2018}
{Rajpurohit}, A.~S., {Allard}, F., {Rajpurohit}, S., {et~al.} 2018, \aap, 620,
  A180

\bibitem[{{Reiners} {et~al.}(2018){Reiners}, {Zechmeister}, {Caballero},
  {Ribas}, {Morales}, {Jeffers}, {Sch{\"o}fer}, {Tal-Or}, {Quirrenbach},
  {Amado}, {Kaminski}, {Seifert}, {Abril}, {Aceituno}, {Alonso-Floriano},
  {Ammler-von Eiff}, {Antona}, {Anglada-Escud{\'e}}, {Anwand-Heerwart},
  {Arroyo-Torres}, {Azzaro}, {Baroch}, {Barrado}, {Bauer}, {Becerril},
  {B{\'e}jar}, {Ben{\'{\i}}tez}, {Berdi{\~n}as}, {Bergond}, {Bl{\"u}mcke},
  {Brinkm{\"o}ller}, {del Burgo}, {Cano}, {C{\'a}rdenas V{\'a}zquez}, {Casal},
  {Cifuentes}, {Claret}, {Colom{\'e}}, {Cort{\'e}s-Contreras}, {Czesla},
  {D{\'{\i}}ez-Alonso}, {Dreizler}, {Feiz}, {Fern{\'a}ndez}, {Ferro},
  {Fuhrmeister}, {Galad{\'{\i}}-Enr{\'{\i}}quez}, {Garcia-Piquer},
  {Garc{\'{\i}}a Vargas}, {Gesa}, {G{\'o}mez Galera}, {Gonz{\'a}lez
  Hern{\'a}ndez}, {Gonz{\'a}lez-Peinado}, {Gr{\"o}zinger}, {Grohnert},
  {Gu{\`a}rdia}, {Guenther}, {Guijarro}, {de Guindos}, {Guti{\'e}rrez-Soto},
  {Hagen}, {Hatzes}, {Hauschildt}, {Hedrosa}, {Helmling}, {Henning}, {Hermelo},
  {Hern{\'a}ndez Arab{\'{\i}}}, {Hern{\'a}ndez Casta{\~n}o}, {Hern{\'a}ndez
  Hernando}, {Herrero}, {Huber}, {Huke}, {Johnson}, {de Juan}, {Kim}, {Klein},
  {Kl{\"u}ter}, {Klutsch}, {K{\"u}rster}, {Lafarga}, {Lamert}, {Lamp{\'o}n},
  {Lara}, {Laun}, {Lemke}, {Lenzen}, {Launhardt}, {L{\'o}pez del Fresno},
  {L{\'o}pez-Gonz{\'a}lez}, {L{\'o}pez-Puertas}, {L{\'o}pez Salas},
  {L{\'o}pez-Santiago}, {Luque}, {Mag{\'a}n Madinabeitia}, {Mall}, {Mancini},
  {Mandel}, {Marfil}, {Mar{\'{\i}}n Molina}, {Maroto Fern{\'a}ndez},
  {Mart{\'{\i}}n}, {Mart{\'{\i}}n-Ruiz}, {Marvin}, {Mathar}, {Mirabet},
  {Montes}, {Moreno-Raya}, {Moya}, {Mundt}, {Nagel}, {Naranjo}, {Nortmann},
  {Nowak}, {Ofir}, {Oreiro}, {Pall{\'e}}, {Panduro}, {Pascual}, {Passegger},
  {Pavlov}, {Pedraz}, {P{\'e}rez-Calpena}, {P{\'e}rez Medialdea}, {Perger},
  {Perryman}, {Pluto}, {Rabaza}, {Ram{\'o}n}, {Rebolo}, {Redondo}, {Reffert},
  {Reinhart}, {Rhode}, {Rix}, {Rodler}, {Rodr{\'{\i}}guez},
  {Rodr{\'{\i}}guez-L{\'o}pez}, {Rodr{\'{\i}}guez Trinidad}, {Rohloff},
  {Rosich}, {Sadegi}, {S{\'a}nchez-Blanco}, {S{\'a}nchez Carrasco},
  {S{\'a}nchez-L{\'o}pez}, {Sanz-Forcada}, {Sarkis}, {Sarmiento},
  {Sch{\"a}fer}, {Schmitt}, {Schiller}, {Schweitzer}, {Solano}, {Stahl},
  {Strachan}, {St{\"u}rmer}, {Su{\'a}rez}, {Tabernero}, {Tala}, {Trifonov},
  {Tulloch}, {Ulbrich}, {Veredas}, {Vico Linares}, {Vilardell}, {Wagner},
  {Winkler}, {Wolthoff}, {Xu}, {Yan}, \& {Zapatero Osorio}}]{Reiners2018a}
{Reiners}, A., {Zechmeister}, M., {Caballero}, J.~A., {et~al.} 2018, \aap, 612,
  A49

\bibitem[{{Rodr{\'\i}guez Mart{\'\i}nez} {et~al.}(2019){Rodr{\'\i}guez
  Mart{\'\i}nez}, {Ballard}, {Mayo}, {Vanderburg}, {Montet}, \&
  {Christiansen}}]{Rodriguez2019}
{Rodr{\'\i}guez Mart{\'\i}nez}, R., {Ballard}, S., {Mayo}, A., {et~al.} 2019,
  \aj, 158, 135

\bibitem[{{Rojas-Ayala} {et~al.}(2010){Rojas-Ayala}, {Covey}, {Muirhead}, \&
  {Lloyd}}]{Rojas-Ayala2010}
{Rojas-Ayala}, B., {Covey}, K.~R., {Muirhead}, P.~S., \& {Lloyd}, J.~P. 2010,
  \apjl, 720, L113

\bibitem[{{Rojas-Ayala} {et~al.}(2012){Rojas-Ayala}, {Covey}, {Muirhead}, \&
  {Lloyd}}]{RojasAyala2012}
{Rojas-Ayala}, B., {Covey}, K.~R., {Muirhead}, P.~S., \& {Lloyd}, J.~P. 2012,
  \apj, 748, 93

\bibitem[{Rosenblatt(1956)}]{Rosenblatt1956}
Rosenblatt, M. 1956, The Annals of Mathematical Statistics, 27, 832

\bibitem[{{Sarro} {et~al.}(2018){Sarro}, {Ordieres-Mer{\'e}},
  {Bello-Garc{\'\i}a}, {Gonz{\'a}lez-Marcos}, \& {Solano}}]{Sarro2018}
{Sarro}, L.~M., {Ordieres-Mer{\'e}}, J., {Bello-Garc{\'\i}a}, A.,
  {Gonz{\'a}lez-Marcos}, A., \& {Solano}, E. 2018, \mnras, 476, 1120

\bibitem[{{Schlaufman} \& {Laughlin}(2010)}]{SchlaufmanLaughlin2010}
{Schlaufman}, K.~C. \& {Laughlin}, G. 2010, \aap, 519, A105

\bibitem[{Schmidhuber(2015)}]{schmidhuber2015deep}
Schmidhuber, J. 2015, Neural networks, 61, 85

\bibitem[{{Schweitzer} {et~al.}(2019){Schweitzer}, {Passegger}, {Cifuentes},
  {B{\'e}jar}, {Cort{\'e}s-Contreras}, {Caballero}, {del Burgo}, {Czesla},
  {K{\"u}rster}, {Montes}, {Zapatero Osorio}, {Ribas}, {Reiners},
  {Quirrenbach}, {Amado}, {Aceituno}, {Anglada-Escud{\'e}}, {Bauer},
  {Dreizler}, {Jeffers}, {Guenther}, {Henning}, {Kaminski}, {Lafarga},
  {Marfil}, {Morales}, {Schmitt}, {Seifert}, {Solano}, {Tabernero}, \&
  {Zechmeister}}]{Schweitzer2019}
{Schweitzer}, A., {Passegger}, V.~M., {Cifuentes}, C., {et~al.} 2019, \aap,
  625, A68

\bibitem[{Scott(1979)}]{Scott1979}
Scott, D.~W. 1979, Biometrika, 66, 605

\bibitem[{{S{\'e}gransan} {et~al.}(2003){S{\'e}gransan}, {Kervella},
  {Forveille}, \& {Queloz}}]{Segransan2003}
{S{\'e}gransan}, D., {Kervella}, P., {Forveille}, T., \& {Queloz}, D. 2003,
  \aap, 397, L5

\bibitem[{Shallue \& Vanderburg(2018)}]{shallue2018identifying}
Shallue, C.~J. \& Vanderburg, A. 2018, The Astronomical Journal, 155, 94

\bibitem[{Sharma {et~al.}(2019)Sharma, Kembhavi, Kembhavi, Sivarani, Abraham,
  \& Vaghmare}]{Sharma2019}
Sharma, K., Kembhavi, A., Kembhavi, A., {et~al.} 2019, Monthly Notices of the
  Royal Astronomical Society, 491, 2280

\bibitem[{{Shetrone} {et~al.}(2015){Shetrone}, {Bizyaev}, {Lawler}, {Allende
  Prieto}, {Johnson}, {Smith}, {Cunha}, {Holtzman}, {Garc{\'\i}a P{\'e}rez},
  {M{\'e}sz{\'a}ros}, {Sobeck}, {Zamora}, {Garc{\'\i}a-Hern{\'a}ndez}, {Souto},
  {Chojnowski}, {Koesterke}, {Majewski}, \& {Zasowski}}]{Shetrone2015}
{Shetrone}, M., {Bizyaev}, D., {Lawler}, J.~E., {et~al.} 2015, \apjs, 221, 24

\bibitem[{{Singh} {et~al.}(1998){Singh}, {Gulati}, \& {Gupta}}]{Singh1998}
{Singh}, H.~P., {Gulati}, R.~K., \& {Gupta}, R. 1998, \mnras, 295, 312

\bibitem[{{Smette} {et~al.}(2015){Smette}, {Sana}, {Noll}, {Horst}, {Kausch},
  {Kimeswenger}, {Barden}, {Szyszka}, {Jones}, {Gallenne}, {Vinther},
  {Ballester}, \& {Taylor}}]{Smette2015}
{Smette}, A., {Sana}, H., {Noll}, S., {et~al.} 2015, \aap, 576, A77

\bibitem[{{Sneden}(1973)}]{Sneden1973}
{Sneden}, C.~A. 1973, PhD thesis, THE UNIVERSITY OF TEXAS AT AUSTIN.

\bibitem[{{Souto} {et~al.}(2017){Souto}, {Cunha}, {Garc{\'\i}a-Hern{\'a}ndez},
  {Zamora}, {Allende Prieto}, {Smith}, {Mahadevan}, {Blake}, {Johnson},
  {J{\"o}nsson}, {Pinsonneault}, {Holtzman}, {Majewski}, {Shetrone}, {Teske},
  {Nidever}, {Schiavon}, {Sobeck}, {Garc{\'\i}a P{\'e}rez}, {G{\'o}mez Maqueo
  Chew}, \& {Stassun}}]{Souto2017}
{Souto}, D., {Cunha}, K., {Garc{\'\i}a-Hern{\'a}ndez}, D.~A., {et~al.} 2017,
  \apj, 835, 239

\bibitem[{{Souto} {et~al.}(2020){Souto}, {Cunha}, {Smith}, {Allende Prieto},
  {Burgasser}, {Covey}, {Garc{\'\i}a-Hern{\'a}ndez}, {Holtzman}, {Johnson},
  {J{\"o}nsson}, {Mahadevan}, {Majewski}, {Masseron}, {Shetrone},
  {Rojas-Ayala}, {Sobeck}, {Stassun}, {Terrien}, {Teske}, {Wanderley}, \&
  {Zamora}}]{Souto2020}
{Souto}, D., {Cunha}, K., {Smith}, V.~V., {et~al.} 2020, \apj, 890, 133

\bibitem[{{Souto} {et~al.}(2018){Souto}, {Unterborn}, {Smith}, {Cunha},
  {Teske}, {Covey}, {Rojas-Ayala}, {Garc{\'\i}a-Hern{\'a}ndez}, {Stassun},
  {Zamora}, {Masseron}, {Johnson}, {Majewski}, {J{\"o}nsson}, {Gilhool},
  {Blake}, \& {Santana}}]{Souto2018}
{Souto}, D., {Unterborn}, C.~T., {Smith}, V.~V., {et~al.} 2018, \apj, 860, L15

\bibitem[{Steels(1993)}]{steels1993artificial}
Steels, L. 1993, Artificial life, 1, 75

\bibitem[{{Tang} {et~al.}(2014){Tang}, {Bressan}, {Rosenfield}, {Slemer},
  {Marigo}, {Girardi}, \& {Bianchi}}]{Tang2014}
{Tang}, J., {Bressan}, A., {Rosenfield}, P., {et~al.} 2014, \mnras, 445, 4287

\bibitem[{{Terrien} {et~al.}(2012){Terrien}, {Mahadevan}, {Bender},
  {Deshpande}, {Ramsey}, \& {Bochanski}}]{Terrien2012}
{Terrien}, R.~C., {Mahadevan}, S., {Bender}, C.~F., {et~al.} 2012, \apjl, 747,
  L38

\bibitem[{Terrien {et~al.}(2015)Terrien, Mahadevan, Bender, Deshpande, \&
  Robertson}]{Terrien2015}
Terrien, R.~C., Mahadevan, S., Bender, C.~F., Deshpande, R., \& Robertson, P.
  2015, The Astrophysical Journal, 802, L10

\bibitem[{{Valenti} \& {Fischer}(2005)}]{ValentiFischer2005}
{Valenti}, J.~A. \& {Fischer}, D.~A. 2005, \apjs, 159, 141

\bibitem[{{Valenti} \& {Piskunov}(1996)}]{ValentiPiskunov1996}
{Valenti}, J.~A. \& {Piskunov}, N. 1996, \aaps, 118, 595

\bibitem[{{Veyette} {et~al.}(2017){Veyette}, {Muirhead}, {Mann}, {Brewer},
  {Allard}, \& {Homeier}}]{Veyette2017}
{Veyette}, M.~J., {Muirhead}, P.~S., {Mann}, A.~W., {et~al.} 2017, \apj, 851,
  26

\bibitem[{{von Braun} {et~al.}(2014){von Braun}, {Boyajian}, {van Belle},
  {Kane}, {Jones}, {Farrington}, {Schaefer}, {Vargas}, {Scott}, {ten
  Brummelaar}, {Kephart}, {Gies}, {Ciardi}, {L{\'o}pez-Morales}, {Mazingue},
  {McAlister}, {Ridgway}, {Goldfinger}, {Turner}, \& {Sturmann}}]{vonBraun2014}
{von Braun}, K., {Boyajian}, T.~S., {van Belle}, G.~T., {et~al.} 2014, \mnras,
  438, 2413

\bibitem[{{von Hippel} {et~al.}(1994){von Hippel}, {Storrie-Lombardi},
  {Storrie-Lombardi}, \& {Irwin}}]{vonHippel1994}
{von Hippel}, T., {Storrie-Lombardi}, L.~J., {Storrie-Lombardi}, M.~C., \&
  {Irwin}, M.~J. 1994, \mnras, 269, 97

\bibitem[{Whitten {et~al.}(2019)Whitten, Placco, Beers, Chies-Santos, Bonatto,
  Varela, Crist{\'{o}}bal-Hornillos, Ederoclite, Masseron, Lee, Akras,
  Fernandes, Caballero, Cenarro, Coelho, Costa-Duarte, Daflon, Dupke,
  de~Oliveira, L{\'{o}}pez-Sanjuan, Mar{\'{\i}}n-Franch, de~Oliveira, Moles,
  Orsi, Rossi, Sodr{\'{e}}, \& Rami{\'{o}}}]{Whitten2019}
Whitten, D.~D., Placco, V.~M., Beers, T.~C., {et~al.} 2019, Astronomy {\&}
  Astrophysics, 622, A182

\bibitem[{{Woolf} \& {Wallerstein}(2006)}]{WoolfWallerstein2006}
{Woolf}, V.~M. \& {Wallerstein}, G. 2006, \pasp, 118, 218

\bibitem[{Wu {et~al.}(2018)Wu, Wong, Rudnick, Shabala, Alger, Banfield, Ong,
  White, Garon, Norris, Andernach, Tate, Lukic, Tang, Schawinski, \&
  Diakogiannis}]{Wu2018}
Wu, C., Wong, O.~I., Rudnick, L., {et~al.} 2018, Monthly Notices of the Royal
  Astronomical Society, 482, 1211

\bibitem[{{Zboril} \& {Byrne}(1998)}]{ZborilByrne1998}
{Zboril}, M. \& {Byrne}, P.~B. 1998, \mnras, 299, 753

\bibitem[{{Zboril} {et~al.}(1998){Zboril}, {Byrne}, \&
  {Rolleston}}]{Zboril1998}
{Zboril}, M., {Byrne}, P.~B., \& {Rolleston}, W.~R.~J. 1998, \mnras, 301, 1104

\bibitem[{{Zechmeister} {et~al.}(2014){Zechmeister}, {Anglada-Escud{\'e}}, \&
  {Reiners}}]{Zechmeister2014}
{Zechmeister}, M., {Anglada-Escud{\'e}}, G., \& {Reiners}, A. 2014, \aap, 561,
  A59

\bibitem[{{Zechmeister} {et~al.}(2018){Zechmeister}, {Reiners}, {Amado},
  {Azzaro}, {Bauer}, {B{\'e}jar}, {Caballero}, {Guenther}, {Hagen}, {Jeffers},
  {Kaminski}, {K{\"u}rster}, {Launhardt}, {Montes}, {Morales}, {Quirrenbach},
  {Reffert}, {Ribas}, {Seifert}, {Tal-Or}, \& {Wolthoff}}]{Zechmeister2018}
{Zechmeister}, M., {Reiners}, A., {Amado}, P.~J., {et~al.} 2018, \aap, 609, A12

\bibitem[{Zhang {et~al.}(1998)Zhang, Patuwo, \& Hu}]{zhang1998forecasting}
Zhang, G., Patuwo, B.~E., \& Hu, M.~Y. 1998, International journal of
  forecasting, 14, 35

\bibitem[{Zheng {et~al.}(2018)Zheng, Wang, \& Ordieres-Mer{\'{e}}}]{Zheng2018}
Zheng, X., Wang, M., \& Ordieres-Mer{\'{e}}, J. 2018, Sensors, 18, 2146

\end{thebibliography}

\newpage
\appendix

\end{document}